\documentclass{article}
\usepackage{fuzz}
\usepackage{url}
\usepackage{listings}

\title{Z Specification for the W3C Editor's Draft Core SHACL Semantics}
\author{Arthur Ryman, {\tt arthur.ryman@gmail.com}}
\date{\today}

\begin{document}
\bibliographystyle{acm}

\maketitle

\begin{abstract}
This article provides a formalization of the W3C Draft Core SHACL Semantics specification using Z notation.
This formalization exercise has identified a number of quality issues in the draft.
It has also established that the recursive definitions in the draft are well-founded.
Further formal validation of the draft will require the use of an executable specification technology.
\end{abstract}

\section{Introduction}
\label{sec-introduction}
The W3C RDF Data Shapes Working Group \cite{w3c:shapeswg} is developing SHACL, a new language for describing constraints on RDF graphs.
A semantics for Core SHACL has been proposed \cite{iovka:core-shacl}, hereafter referred to as the {\em semantics draft}.
The proposed semantics includes an abstract syntax, inference rules, and a definition of typing which allows for certain kinds of recursion.
The semantics draft uses precise mathematical language, but is informal in the sense that it is not written in a formal specification language
and therefore cannot benefit from tools such as type-checkers.

This document provides a formal translation of the semantics draft into Z Notation  \cite{spivey:zrm}.
The \LaTeX\ source for this article has been type-checked using the \fuzz\ type-checker \cite{spivey:fuzz}
and is available in the GitHub repository  \cite{agryman:z-core-shacl-semantics}
{\tt agryman/z-core-shacl-semantics}.

Our motive for formalizing and type-checking the semantics draft is to help to improve its quality
and the ultimate design of SHACL. 

\subsection{Organization of this Article}
The remainder of this article is organized as follows.
\begin{itemize}
\item Section~\ref{sec-basic-rdf-concepts} formalizes some basic RDF concepts.
\item Section~\ref{sec-abstract-syntax} translates the abstract syntax of SHACL into Z notation.
\item Section~\ref{sec-evaluation} formalizes the evaluation semantics of SHACL.
\item Section~\ref{sec-declarative-semantics} formalizes the declarative semantics of shape expression schemas.
\item Section~\ref{sec-issues} summarizes the quality issues found in the draft.
\item Section~\ref{sec-conclusion} concludes with some remarks about the benefits of the formalization exercise and 
possible next steps.
\end{itemize}

\section{Basic RDF Concepts}
\label{sec-basic-rdf-concepts}
This section formalizes some basic RDF concepts.
We reuse some formal definitions given in \cite{arthur:recursion}, modifying the identifiers to match those used in the semantics draft.

\subsection{$TERM$}
Let $TERM$ be the set of all RDF {\em terms}.
\begin{zed}
	[TERM]
\end{zed}

\subsection{$Iri$, $Blank$, and $Lit$}
The set of all RDF terms is partitioned into IRIs, blank nodes, and literals.
\begin{axdef}
	Iri, Blank, Lit: \power TERM
\where
	\langle Iri, Blank, Lit \rangle \partition TERM
\end{axdef}

\subsection{$IRI$}
The semantics draft introduces the term $Iri$, but it uses the term $IRI$ in the definitions of the abstract syntax.
We treat $IRI$ as a synonym for $Iri$.
\begin{zed}
	IRI == Iri
\end{zed}

\subsection{$Triple$}
An RDF triple is an ordered triple of RDF terms referred to as the subject, predicate, and object.
\begin{zed}
	Triple == \{~ s, p, o:TERM | s \notin Lit \land p \in IRI ~\}
\end{zed}
\begin{itemize}
\item The subject is not a literal.
\item The predicate is an IRI.
\end{itemize}

\subsection{$subject$, $predicate$, and $object$}
It is convenient to define generic functions that select the first, second, or third component of a Cartesian product of three sets.
\begin{zed}
	fst[X,Y,Z] == (\lambda x:X; y:Y; z:Z @ x~)
\also
	snd[X,Y,Z] == (\lambda x:X; y:Y; z:Z @ y~)
\also
	trd[X,Y,Z] == (\lambda x:X; y:Y; z:Z @ z~)
\end{zed}

The subject, predicate, and object of an RDF triple are the terms that appear in the corresponding positions.
\begin{zed}
	subject == (\lambda t:Triple @ fst(t) ~)
\also
	predicate == (\lambda t:Triple @ snd(t) ~)
\also
	object == (\lambda t:Triple @ trd(t) ~)
\end{zed}

\subsection{$Graph$}
An RDF graph is a finite set of RDF triples.
\begin{zed}
	Graph == \finset Triple
\end{zed}

\subsection{$subjects$, $predicates$, and $objects$}
The subjects, predicates, and objects of a graph are the sets of RDF terms that appear in the corresponding positions of its triples.
\begin{zed}
	subjects == (\lambda g: Graph @ \{~ t:g @ subject(t) ~\} ~)
\also
	predicates == (\lambda g: Graph @ \{~ t:g @ predicate(t) ~\} ~)
\also
	objects == (\lambda g: Graph @ \{~ t:g @ object(t) ~\} ~)
\end{zed}

\subsection{$nodes$}
The nodes of an RDF are its subjects and objects.
\begin{zed}
	nodes == (\lambda g: Graph @ subjects(g) \cup objects(g) ~)
\end{zed}

\subsection{$PointedGraph$}
A pointed graph is a graph and a distinguished node in the graph.
The distinguished node is variously referred to as the start, base, or focus node of the pointed graph, depending on the context.
\begin{zed}
	PointedGraph == \{~ g: Graph; n: TERM | n \in nodes(g) ~\}
\end{zed}

\section{Abstract Syntax}
\label{sec-abstract-syntax}
This section contains a translation of the abstract syntax of SHACL into Z.
The semantics draft defines the abstract syntax using an informal Extended Backus-Naur Form (EBNF).

The approach used here is to interpret each term or expression that appears in the abstract syntax as
a mathematical set that is isomorphic to the set of abstract syntax tree fragments denoted by the corresponding term or expression.
Care has been taken to preserve the exact spelling and case of each abstract syntax term so that there is a direct correspondence
between the abstract syntax and Z.
For example, the term {\tt Schema} is interpreted as the set $Schema$.

We give a Z definition for each abstract syntax term that appears on the left-hand side of the EBNF definition operator ({\tt ::=}).
The order in which these terms appear in the semantics draft has been preserved in this document.
If a Z term has a corresponding EBNF rule, we include it here for easy reference.
Refer to \cite{iovka:core-shacl} for the complete definition of the abstract syntax.

A sequence of two or more abstract syntax terms is interpreted as the Cartesian product of the corresponding sets, i.e.
{\tt A B} is interpreted as $A \cross B$.

The abstract syntax Kleene star ({\tt *}) and plus ({\tt +}) operators are interpreted as sequence ($\seq$) 
and non-empty sequence ($\seq_1$) operators on the corresponding sets, i.e.
{\tt A+} is interpreted as $\seq_1 A$.

The abstract syntax optional operator ({\tt ?}) is interpreted as taking the union of the set of singletons and the 
empty set of the corresponding set
using the generic function $OPTIONAL$ (defined below), i.e.
{\tt A?} is interpreted as $OPTIONAL[A]$.

Abstract syntax terms that are defined as alternations ({\tt |}) of two or more expressions are translated into either free types or unions
of sets.
A side effect of this process is that constructors may be required for each branch of the alternation.
In some cases the name of the constructors can be derived from a corresponding element of the abstract syntax.
For example, in {\tt ShapeDefinition}, {\tt open} and {\tt close} are mapped to the constructors $open$ and $close$.
In the cases where there is no convenient element of the abstract syntax, we mint new constructor names.

We also introduce new Z identifiers when an element of the abstract syntax does not map to a valid alphanumeric Z identifier.
For example the the shape label negation operator ({\tt !}) is mapped to $negate$.

\subsection{$OPTIONAL$}
An optional value is represented by a singleton set, if the value is present, or the empty set, if the value is absent.
\begin{zed}
	OPTIONAL[X] == \{~v:X@\{v\}~\} \cup \{ \emptyset \}
\end{zed}

\subsection{$Schema$}

\begin{verbatim}
Schema ::= Rule+
\end{verbatim}

A schema is a sequence of one or more rules.
\begin{zed}
	Schema == \seq_1 Rule
\end{zed}

\subsection{$Rule$}

\begin{verbatim}
Rule ::= ShapeLabel ShapeDefinition ExtensionCondition*
\end{verbatim}

A rule consists of a shape label, a shape definition, and a sequence of zero or more extension conditions.
\begin{zed}
	Rule == ShapeLabel \cross ShapeDefinition \cross \seq ExtensionCondition
\end{zed}

It is convenient to introduce functions that select the components of a rule.
\begin{zed}
	shapeLabel == (\lambda r:Rule @ fst(r) ~)
\also
	shapeDef == (\lambda r:Rule @ snd(r) ~)
\also
	extConds == (\lambda r:Rule @ trd(r) ~)
\end{zed}

\subsection{$ShapeLabel$}

\begin{verbatim}
ShapeLabel ::= an identifier
\end{verbatim}

A shape label is an identifier drawn from some given set.
\begin{zed}
	[ShapeLabel]
\end{zed}

\subsection{$ShapeDefinition$}

\begin{verbatim}
ShapeDefinition ::= ClosedShape | OpenShape
\end{verbatim}

A shape definition is either a closed shape or an open shape.
\begin{zed}
	ShapeDefinition ::= \\
\t1		close \ldata ShapeExpr \rdata | \\
\t1		open \ldata OPTIONAL[InclPropSet] \cross ShapeExpr \rdata
\end{zed}
Note that abstract syntax terms that are defined using alternation are naturally represented as free types in Z Notation.
\begin{itemize}
\item $close$ is the constructor for closed shapes.A closed shape consists of a shape expression.
\item $open$ is the constructor for open shapes. An open shape consists of an optional included properties set and a shape expression.
\end{itemize}

Given a shape definition $d$, let $shapeExpr(d)$ be its shape expression.
\begin{axdef}
	shapeExpr: ShapeDefinition \fun ShapeExpr
\where
	\forall x: ShapeExpr @ \\
\t1		shapeExpr(close(x)) = x
\also
	\forall o: OPTIONAL[InclPropSet]; x: ShapeExpr @ \\
\t1		shapeExpr(open(o,x)) = x
\end{axdef}

\subsection{$ClosedShape$}

\begin{verbatim}
ClosedShape ::= 'close' ShapeExpr
\end{verbatim}

The set of closed shapes is the range of the $close$ shape definition constructor.
\begin{zed}
	ClosedShape == \ran close
\end{zed}

\subsection{$OpenShape$}

\begin{verbatim}
OpenShape ::= 'open' InclPropSet? ShapeExpr
\end{verbatim}

The set of open shapes is the range of the $open$ shape definition constructor.
\begin{zed}
	OpenShape == \ran open
\end{zed}

\subsection{$InclPropSet$}

\begin{verbatim}
InclPropSet ::= PropertiesSet
\end{verbatim}

An included properties set is a properties set.
\begin{zed}
	InclPropSet == PropertiesSet
\end{zed}
Note that there seems little motivation to introduce the term $InclPropSet$ since it is identical to $PropertiesSet$.

\subsection{$PropertiesSet$}

\begin{verbatim}
PropertiesSet ::= set of IRI
\end{verbatim}

A properties set is a set of IRIs.
\begin{zed}
	PropertiesSet == \power IRI
\end{zed}

\subsection{$ShapeExpr$}

\begin{verbatim}
ShapeExpr ::= EmptyShape
| TripleConstraint Cardinality
| InverseTripleConstraint Cardinality
| NegatedTripleConstraint
| NegatedInverseTripleConstraint
| SomeOfShape
| OneOfShape
| GroupShape
| RepetitionShape
\end{verbatim}

A shape expression defines constraints on RDF graphs.
\begin{zed}
	ShapeExpr ::= \\
\t1		emptyshape | \\
\t1		triple \ldata DirectedTripleConstraint \cross Cardinality \rdata | \\
\t1		someOf \ldata \seq_1 ShapeExpr \rdata | \\
\t1		oneOf \ldata \seq_1 ShapeExpr \rdata | \\
\t1		group \ldata \seq_1 ShapeExpr \rdata | \\
\t1		repetition \ldata ShapeExpr \cross Cardinality \rdata
\end{zed}
\begin{itemize}
\item $emptyshape$ is the empty shape expression.
\item $triple$ is the constructor for triple constraint shape expressions. 
A triple constraint shape expression consists of a directed triple constraint and a cardinality.
\item $someOf$ is the constructor for some-of shape expressions.
A some-of shape expression consists of a sequence of one or more shape expressions.
\item $oneOf$ is the constructor for one-of shape expressions.
A one-of shape expression consists of a sequence of one or more shape expressions.
\item $group$ is the constructor for grouping shape expressions.
A grouping shape expression consists of a sequence of one or more shape expressions.
\item $repetition$ is the constructor for repetition shape expressions.
A repetition shape expression consists of a shape expression and a cardinality.
\end{itemize}

\subsection{$EmptyShape$}

\begin{verbatim}
EmptyShape ::= 'emptyshape'
\end{verbatim}

The set of empty shape expressions is the singleton set that contains the empty shape.
\begin{zed}
	EmptyShape == \{ emptyshape \}
\end{zed}

\subsection{$DirectedPredicate$}

A directed predicate is an IRI with a direction that indicates its usage in a triple. 
$nop$ indicates the normal direction, namely the predicate relates the subject node to the object node. 
$inv$ indicates the inverse direction, namely the predicate relates the object node to the subject node.
\begin{zed}
	DirectedPredicate ::= \\
\t1		nop \ldata IRI \rdata | \\
\t1		inv \ldata IRI \rdata
\end{zed}

The semantics draft uses the notation {\tt \verb+^+p} for $inv(p)$.

Let $predDF(dp)$ denote the predicate of a directed predicate $dp$.
\begin{axdef}
	predDP : DirectedPredicate \fun IRI
\where
	\forall p: IRI @ \\
\t1		predDP(nop(p)) = predDP(inv(p)) = p
\end{axdef}

\subsection{$DirectedTripleConstraint$}
A directed triple constraint consists of a directed predicate and a constraint.
The constraint is a value or shape constraint on the object of a triple if the direction is normal, 
or a shape constraint on the subject of a triple if the direction is inverted.
\begin{zed}
	DirectedTripleConstraint == \\
\t1		\{~ dp: DirectedPredicate; C: Constraint | \\
\t2			dp \in \ran inv \implies C \in ShapeConstr ~\}
\end{zed}

The semantics draft uses the notation {\tt p::C} for $(nop(p),C)$ and {\tt \verb+^+p::C} for $(inv(p),C)$.

Let $predDTC(dtc)$ denote the predicate of the directed triple constraint $dtc$.
\begin{axdef}
	predDTC: DirectedTripleConstraint \fun IRI
\where
	\forall dp: DirectedPredicate; C: Constraint | \\
\t1		(dp, C) \in DirectedTripleConstraint @ \\
\t2			predDTC(dp, C) = predDP(dp)
\end{axdef}

Let $constrDTC(dtc)$ denote the constraint of the directed triple constraint $dtc$.
\begin{axdef}
	constrDTC: DirectedTripleConstraint \fun Constraint
\where
	\forall dp: DirectedPredicate; C: Constraint | \\
\t1		(dp, C) \in DirectedTripleConstraint @ \\
\t2			constrDTC(dp, C) = C
\end{axdef}

\subsection{$TripleConstraint$}

\begin{verbatim}
TripleConstraint ::= IRI ValueConstr | IRI ShapeConstr
\end{verbatim}

A triple constraint places conditions on triples whose subject is a given focus node
and whose predicate is a given IRI.
\begin{axdef}
	TripleConstraint: \power DirectedTripleConstraint
\where
	TripleConstraint = \\
\t1		\{~ p: IRI; C: Constraint @ (nop(p),C) ~\}
\end{axdef}

\subsection{$InverseTripleConstraint$}

\begin{verbatim}
InverseTripleConstraint ::= '^' IRI ShapeConstr
\end{verbatim}

An inverse triple constraint places conditions on triples whose object is a given focus node
and whose predicate is a given IRI.
\begin{axdef}
	InverseTripleConstraint: \power DirectedTripleConstraint
\where
	InverseTripleConstraint = \\
\t1		\{~ p: IRI; C: ShapeConstr @ (inv(p),C) ~\}
\end{axdef}

\subsection{$Constraint$}
A constraint is a condition on the object node of a triple for normal predicates 
or the subject node of a triple for inverse predicates.
\begin{zed}
	Constraint ::= \\
\t1		valueSet \ldata \power (Lit \cup IRI) \rdata | \\
\t1		datatype \ldata LiteralDatatype \cross OPTIONAL[XSFacet] \rdata | \\
\t1		kind \ldata NodeKind \rdata | \\
\t1		or \ldata \seq_1 ShapeLabel \rdata | \\
\t1		and \ldata \seq_1 ShapeLabel \rdata | \\
\t1		nor \ldata \seq_1 ShapeLabel \rdata | \\
\t1		nand \ldata \seq_1 ShapeLabel \rdata
\end{zed}
\begin{itemize}
\item $valueSet$ is the constructor for value set value constraints.
A value set value constraint consists of a set of literals and IRIs.
\item $datatype$ is the constructor for literal datatype value constraints.
A literal datatype value constraint consists of a literal datatype and an optional XML Schema facet.
\item $kind$ is the constructor for node kind value constraints.
A node kind value constraint consists of a specification for a subset of RDF terms.
\item $or$ is the constructor for disjunction shape constraints.
A node must satisfy at least one of the shapes.
\item $and$ is the constructor for conjunction shape constraints.
A node must satisfy all of the shapes.
\item $nor$ is the constructor for negated disjunction shape constraints.
A node must not satisfy any of the shapes.
\item $nand$ is the constructor for negated conjunction shape constraints.
A node must not satisfy all of the shapes.
\end{itemize}

\subsection{$Cardinality$}

\begin{verbatim}
Cardinality ::= '[' MinCardinality ';' MaxCardinality ']'
\end{verbatim}

Cardinality defines a range for the number of elements in a set.
\begin{zed}
	Cardinality == MinCardinality \cross MaxCardinality
\end{zed}
\begin{itemize}
\item A cardinality consists of a minimum cardinality and a maximum cardinality.
\end{itemize}

\subsection{$MinCardinality$}

\begin{verbatim}
MinCardinality ::= a natural number
\end{verbatim}

Minimum cardinality is the minimum number of elements required to be in a set.
\begin{zed}
	MinCardinality == \nat
\end{zed}

\subsection{$MaxCardinality$}

\begin{verbatim}
MaxCardinality ::= a natural number | 'unbound' 
\end{verbatim}

Maximum cardinality is the maximum number of elements required to be in a set.
\begin{zed}
	MaxCardinality ::= maxCard \ldata \nat \rdata | unbound
\end{zed}
\begin{itemize}
\item $maxCard$ is the constructor for finite maximum cardinalities.
A finite maximum cardinality is a natural number.
Note that a maximum cardinality of $0$ means that the set must be empty.
\item $unbound$ indicates that the maximum number of elements in a set is unbounded.
\end{itemize}

\subsection{$inBounds$}
A natural number $k$ is said to be in bounds of a cardinality when $k$ is between the minimum and maximum
limits of the cardinality.
\begin{axdef}
	inBounds: \nat \rel Cardinality
\where
	\forall k, n: \nat @ \\
\t1		k \inrel{inBounds} (n,unbound) \iff n \leq k
\also
	\forall k, n, m: \nat @ \\
\t1		k \inrel{inBounds} (n,maxCard(m)) \iff n \leq k \leq m
\end{axdef}

\subsection{Notation}
Let $a$ be an IRI, let $C$ be a value or shape constraint, let $n$ and $m$ be non-negative integers.
The semantics draft uses the notation listed in Table~\ref{notation-meaning} for some shape expressions.

\begin{table}[h]
\begin{center}
\begin{tabular}{|c|c|}
\hline
Notation				& Meaning \\
\hline
{\tt a::C[n;m]}			& $triple(nop(a,C),(n,maxCard(m)))$\\
{\tt \verb+^+a::C[n;m]}	& $triple(inv(a,C),(n,maxCard(m)))$ \\
\hline
{\tt a::C}				& {\tt a::C[1;1]}\\
{\tt \verb+^+a::C}		& {\tt \verb+^+a::C[1;1]} \\
\hline
{\tt !a::C}				& {\tt a::C[0;0]} \\
{\tt !\verb+^+a:C}		& {\tt \verb+^+a::C[0;0]} \\
\hline
\end{tabular}
\end{center}
\caption{Meaning of shape expression notation}
\label{notation-meaning}
\end{table}
\begin{itemize}
\item If the cardinality is {\tt [1;1]} it may be omitted.
\item The negated shape expressions are semantically equivalent to the corresponding non-negated shape expressions with cardinality {\tt [0;0]}.
\end{itemize}

\subsection{$none$, $one$}
It is convenient to define some common cardinalities.
\begin{zed}
	none == (0,maxCard(0))
\also
	one == (1,maxCard(1))
\end{zed}
\begin{itemize}
\item A cardinality of $none =$ {\tt [0;0]} is used to indicate a negated triple or inverse triple constraint.
\item A cardinality of $one =$ {\tt [1;1]} is the default cardinality of a triple or inverse triple constraint
when no cardinality is explicitly given in the notations {\tt a::C} and {\tt \verb+^+a::C}.
\end{itemize}

\subsection{$NegatedTripleConstraint$}

\begin{verbatim}
NegatedTripleConstraint ::= '!' TripleConstraint
\end{verbatim}

A negated triple constraint shape expression is a triple constraint shape expression that has a cardinality of $none$.
\begin{zed}
	NegatedTripleConstraint == \\
\t1		\{~ tc: TripleConstraint @ triple(tc,none) ~\}
\end{zed}

\subsection{$NegatedInverseTripleConstraint$}

\begin{verbatim}
NegatedInverseTripleConstraint ::= '!' InverseTripleConstraint
\end{verbatim}

A negated inverse triple constraint shape expression is an inverse triple constraint shape expression that has a cardinality of $none$.
\begin{zed}
	NegatedInverseTripleConstraint == \\
\t1		\{~ itc: InverseTripleConstraint @ triple(itc,none) ~\}
\end{zed}

\subsection{$ValueConstr$}

\begin{verbatim}
ValueConstr ::= ValueSet | LiteralDatatype XSFacet? | NodeKind
\end{verbatim}

A value constraint places conditions on the object nodes of triples for normal predicates
and on the subject nodes of triples for inverse predicates.
\begin{zed}
	ValueConstr == \ran valueSet \cup \ran datatype \cup \ran kind
\end{zed}

\subsection{$ValueSet$}

\begin{verbatim}
ValueSet ::= set of literals and IRI
\end{verbatim}

The set of value set value constraints is the range of the $valueSet$ constructor.
\begin{zed}
	ValueSet == \ran valueSet
\end{zed}

\subsection{$LiteralDatatype$}

\begin{verbatim}
LiteralDatatype ::= an RDF literal datatype
\end{verbatim}

A literal datatype is an IRI that identifies a set of literal RDF terms.
We assume that this subset of IRIs is given.
\begin{axdef}
	LiteralDatatype: \power IRI
\end{axdef}

We also assume that we are given an interpretation of each literal datatype as a set of literals.
\begin{axdef}
	literalsOfDatatype: LiteralDatatype \fun \power Lit
\end{axdef}

\subsection{$NodeKind$}

\begin{verbatim}
NodeKind ::= 'iri' | 'blank' | 'literal' | 'nonliteral'
\end{verbatim}

A node kind identifies a subset of RDF terms.
\begin{zed}
	NodeKind ::= iri | blank | literal | nonliteral
\end{zed}
\begin{itemize}
\item $iri$ identifies the set of IRIs.
\item $blank$ identifies the set of blank nodes.
\item $literal$ identifies the set of literals.
\item $nonliteral$ identifies the complement of the set of literals, i.e. the union of IRIs and blank nodes.
\end{itemize}

Each node kind corresponds to a set of RDF terms.
\begin{axdef}
	termsOfKind: NodeKind \fun \power TERM
\where
	termsOfKind(iri) = IRI
\also
	termsOfKind(blank) = Blank
\also
	termsOfKind(literal) = Lit
\also
	termsOfKind(nonliteral) = TERM \setminus Lit
\end{axdef}

\subsection{$XSFacet$}

\begin{verbatim}
XSFacet ::= an XSD restriction
\end{verbatim}

An XML Schema facet places restrictions on literals.
We assume this is a given set.
\begin{zed}
	[XSFacet]
\end{zed}

We also assume that we are given an interpretation of facets as sets of literals.
\begin{axdef}
	literalsOfFacet: LiteralDatatype \cross XSFacet \fun \power Lit
\where
	\forall d: LiteralDatatype; f: XSFacet @ \\
\t1		literalsOfFacet(d,f) \subseteq literalsOfDatatype(d)
\end{axdef}
\begin{itemize}
\item The literals that correspond to a facet of a datatype are a subset of the literals that correspond to the datatype.
\end{itemize}

\subsection{$ShapeConstr$}

\begin{verbatim}
ShapeConstr ::= ('!')? DisjShapeConstr | ConjShapeConstraint
\end{verbatim}

A shape constraint requires that a node satisfy logical combinations of one or more other shapes which are identified by their shape labels.
\begin{zed}
	ShapeConstr == \ran or \cup \ran and \cup \ran nor \cup \ran nand
\end{zed}

\subsection{$DisjShapeConstr$}

\begin{verbatim}
DisjShapeConstr ::= ShapeLabel ('or' ShapeLabel)*
\end{verbatim}

The set of all disjunctive shape constraints is the range of the $or$ constructor.
\begin{zed}
	DisjShapeConstr == \ran or
\end{zed}

\subsection{$ConjShapeConstraint$}

\begin{verbatim}
ConjShapeConstraint ::= ShapeLabel ('and' ShapeLabel)*
\end{verbatim}

The set of all conjunctive shape constraints is the range of the $and$ constructor.
\begin{zed}
	ConjShapeConstraint == \ran and
\end{zed}

\subsection{$SomeOfShape$}

\begin{verbatim}
SomeOfShape ::= ShapeExpr ('|' ShapeExpr)*
\end{verbatim}

The set of some-of shape expressions is the range of $someOf$.
\begin{zed}
	SomeOfShape == \ran someOf
\end{zed}

\subsection{$OneOfShape$}

\begin{verbatim}
OneOfShape ::= ShapeExpr ('@' ShapeExpr)* 
\end{verbatim}

The set of one-of shape expressions is the range of $oneOf$.
\begin{zed}
	OneOfShape == \ran oneOf
\end{zed}

\subsection{$GroupShape$}

\begin{verbatim}
GroupShape ::= ShapeExpr (',' ShapeExpr)*
\end{verbatim}

The set of grouping shape expressions is the range of $group$.
\begin{zed}
	GroupShape == \ran group
\end{zed}

\subsection{$RepetitionShape$}

\begin{verbatim}
RepetitionShape ::= ShapeExpr Cardinality
\end{verbatim}

The set of repetition shape expressions is the range of $repetition$.
\begin{zed}
	RepetitionShape == \ran repetition
\end{zed}

\subsection{$ExtensionCondition$}

\begin{verbatim}
ExtensionCondition ::= ExtLangName ExtDefinition
\end{verbatim}

An extension condition is the definition of a constraint written in an extension language
\begin{zed}
	ExtensionCondition == ExtLangName \cross ExtDefinition
\end{zed}

\subsection{$ExtLangName$}

\begin{verbatim}
ExtLangName ::= an identifier
\end{verbatim}

An extension language name is an identifier for an extension language, such as JavaScript.
We assume this is a given set.
\begin{zed}
	[ExtLangName]
\end{zed}

\subsection{$ExtDefinition$}

\begin{verbatim}
ExtDefinition ::= a string
\end{verbatim}

An extension definition is a program written in some extension language that implements a constraint check.
We assume this is a given set.
\begin{zed}
	[ExtDefinition]
\end{zed}

An extension condition represents a function that takes as input a pointed graph, and returns
as output a boolean with the value {\tt true} if the constraint is violated and {\tt false} is satisfied.
We assume we are given a mapping that associates each extension condition with the set of pointed graphs that violate it.
\begin{axdef}
	violatedBy: ExtensionCondition \fun \power PointedGraph
\end{axdef}

\subsection{$ShapeLabel$ Definitions}
Given a schema $S$, let $defs(S)$ be the set of all shape labels defined in $S$.
\begin{zed}
	defs == (\lambda S: Schema @ \\
\t1		\{~ r: \ran S @ shapeLabel(r) ~\} ~)
\end{zed}
 
Each rule in a schema must be identified by a unique shape label.
\begin{zed}
	SchemaUL == \{~ S: Schema | \# S = \# (defs(S)) ~\}
\end{zed}
\begin{itemize}
\item In a schema with unique rule labels there are as many rules as labels.
\end{itemize}

\subsection{$rule$}
Given a schema $S$ with unique rule labels, and a label $T$ defined in $S$, let $rule(T,S)$ be the corresponding rule.
\begin{axdef}
	rule: ShapeLabel \cross SchemaUL \pfun Rule
\where
	\dom rule = \{~ T: ShapeLabel; S: SchemaUL | T \in defs(S) ~\}
\also
	\forall S: SchemaUL@\\
\t1		\forall r: \ran(S) @ \\
\t2			\LET T == shapeLabel(r) @ \\
\t3				rule(T,S) = r
\end{axdef}

\subsection{$ShapeLabel$ References}
Given a schema $S$, let $refs(S)$ be the set of shape labels referenced in $S$.
\begin{zed}
	refs == (\lambda S: Schema @ \bigcup \{~ r: \ran S @ refsRule(r) ~\} ~)
\end{zed}
\begin{itemize}
\item The set of references in a schema is the union of the sets of references in its rules.
\end{itemize}

Given a rule $r$, let $refsRule(r)$ be the set of shape labels referenced in $r$.
\begin{zed}
	refsRule == (\lambda r: Rule @ refsShapeDefinition(shapeDef(r)) ~)
\end{zed}
\begin{itemize}
\item The set of references in a rule is the set of references in its shape definition.
\end{itemize}

Given a shape definition $d$, let $refsShapeDefinition(d)$ be the set of shape labels referenced in $d$.
\begin{axdef}
	refsShapeDefinition: ShapeDefinition \fun \finset ShapeLabel
\where
	\forall d: ShapeDefinition @ \\
\t1		refsShapeDefinition(d) = refsShapeExpr(shapeExpr(d))
\end{axdef}
\begin{itemize}
\item The set of references in a shape definition is the set of references in its shape expression.
\end{itemize}

Given a shape expression $x$, let $refsShapeExpr(x)$ be the set of shape labels referenced in $x$.
\begin{axdef}
	refsShapeExpr: ShapeExpr \fun \finset ShapeLabel
\where
	refsShapeExpr(emptyshape) = \emptyset
\also
	\forall dtc: DirectedTripleConstraint; c: Cardinality @ \\
\t1		refsShapeExpr(triple(dtc,c)) = \\
\t2			refsDirectedTripleConstraint(dtc)
\also
	\forall xs: \seq_1 ShapeExpr @ \\
\t1		refsShapeExpr(someOf(xs)) = \\
\t1		refsShapeExpr(oneOf(xs)) = \\
\t1		refsShapeExpr(group(xs)) = \\
\t2			\bigcup \{~ x: \ran xs @ refsShapeExpr(x) ~\}
\also
	\forall x: ShapeExpr; c: Cardinality @ \\
\t1		refsShapeExpr(repetition(x,c)) = \\
\t2			refsShapeExpr(x)
\end{axdef}
\begin{itemize}
\item The empty shape expression references no labels.
\item A directed triple constraint shape expression references the labels referenced in the directed triple constraint.
\item A some-of or one-of or group shape expression references the union of the labels referenced in each component shape expression.
\item A repetition shape expression references the labels referenced in its unrepeated shape expression.
\end{itemize}

Given a directed triple constraint $dtc$, let $refsDirectedTripleConstraint(dtc)$ be the set of shape labels referenced in $dtc$.
\begin{axdef}
	refsDirectedTripleConstraint: \\
\t1		DirectedTripleConstraint \fun \finset ShapeLabel
\where
	\forall a: IRI; C: ValueConstr @ \\
\t1		refsDirectedTripleConstraint((nop(a),C)) = \emptyset
\also
	\forall a: IRI; C: ShapeConstr @ \\
\t1		refsDirectedTripleConstraint((nop(a),C)) = \\
\t1		refsDirectedTripleConstraint((inv(a),C)) = \\
\t2			refsShapeConstr(C)
\end{axdef}
\begin{itemize}
\item A value triple constraint references no labels.
\item A shape triple constraint references the labels in its shape constraint.
\end{itemize}

Given a shape constraint $C$, let $refsShapeConstr(C)$ be the set of shape labels referenced in $C$.
\begin{axdef}
	refsShapeConstr: ShapeConstr \fun \finset ShapeLabel
\where
	\forall ls: \seq_1 ShapeLabel @ \\
\t1		refsShapeConstr(or(ls)) = \\
\t1		refsShapeConstr(and(ls)) = \\
\t1		refsShapeConstr(nor(ls)) = \\
\t1		refsShapeConstr(nand(ls)) = \\
\t2			\ran ls
\end{axdef}
\begin{itemize}
\item A shape constraint references the range of its sequence of shape labels.
\end{itemize}

Every shape label referenced in a schema must be defined in the schema.
\begin{zed}
	SchemaRD == \{~ s:Schema | refs(s) \subseteq defs(s) ~\}
\end{zed}

A schema is well-formed if its rules have unique labels and all referenced shape labels are defined.
\begin{zed}
	SchemaWF == SchemaUL \cap SchemaRD
\end{zed}

\section{Evaluation}
\label{sec-evaluation}
This section defines the interpretation of shapes as constraints on RDF graphs.
All functions that are defined in the semantics draft are given formal definitions here.
We assume that from this point on whenever the semantics draft refers to schemas they are well-formed.

\subsection{$shapes$}
Given a well-formed schema $S$, let $shapes(S)$ be the set of shape labels that appear in $S$.
\begin{zed}
	shapes == (\lambda S: SchemaWF @ defs(S) ~)
\end{zed}

\subsection{$expr$}
Given a shape label $T$ and a well-formed schema $S$, let $expr(T,S)$ be the shape expression in the rule with label $T$ in $S$.
\begin{axdef}
	expr: ShapeLabel \cross SchemaWF \pfun ShapeExpr
\where
	\dom expr = \{~ T: ShapeLabel; S: SchemaWF | T \in shapes(S) ~\} \\
\also
	\forall T: ShapeLabel; S: SchemaWF | T \in shapes(S) @ \\
\t1		\LET r == rule(T,S) @ \\
\t2			expr(T,S) = shapeExpr(shapeDef(r))
\end{axdef}
\begin{itemize}
\item The shape expression for a shape label $T$ is the shape expression in the shape definition of the rule $r$ that has shape label $T$.
\end{itemize}

\subsection{$incl$}

Given a shape label $T$ defined in a well-formed schema $S$, let $incl(T,S)$ be the,
possibly empty, set of included properties.
\begin{axdef}
	incl: ShapeLabel \cross SchemaWF \pfun InclPropSet
\where
	\dom incl = \{~ T: ShapeLabel; S: SchemaWF | T \in shapes(S) ~\}
\also
	\forall T: ShapeLabel; S: SchemaWF | T \in shapes(S) @ \\
\t1		\exists_1 r: \ran S | T = shapeLabel(r) @ \\
\t2			incl(T,S) = inclShapeDefinition(shapeDef(r))
\end{axdef}
\begin{itemize}
\item The included properties set for a shape label $T$ is the included properties set in the shape definition of the rule $r$ that has shape label $T$.
\end{itemize}

Given a shape definition $d$, let $inclShapeDefinition(d)$ be its included properties set.
\begin{axdef}
	inclShapeDefinition: ShapeDefinition \fun InclPropSet
\where
	\forall x: ShapeExpr @ \\
\t1		inclShapeDefinition(close(x)) = \\
\t1		inclShapeDefinition(open(\{ \emptyset \},x)) \\
\t2			= \emptyset
\also
	\forall ips: InclPropSet; x: ShapeExpr @ \\
\t1		inclShapeDefinition(open(\{ ips \},x)) = ips
\end{axdef}
\begin{itemize}
\item The included property set of a closed shape definition or an open definition with no included
property set is the empty set.
\item The included property set of an open shape definition with an included property set is that
included property set.
\end{itemize}

\subsection{$properties$}

Given a shape expression $x$, let $properties(x)$ be the set of properties that appear in some 
triple constraint in $x$.
\begin{axdef}
	properties: ShapeExpr \fun PropertiesSet
\where
	properties(emptyshape) = \emptyset
\also
	\forall tc: TripleConstraint; c: Cardinality @ \\
\t1		properties(triple(tc,c)) = \\
\t2			propertiesTripleConstraint(tc)
\also
	\forall itc: InverseTripleConstraint; c: Cardinality @ \\
\t1		properties(triple(itc,c)) = \\
\t2			\emptyset
\also
	\forall xs: \seq_1 ShapeExpr @ \\
\t1		properties(someOf(xs)) = \\
\t1		properties(oneOf(xs)) = \\
\t1		properties(group(xs)) = \\
\t2			\bigcup \{~ x: \ran xs @ properties(x) ~\}
\also
	\forall x: ShapeExpr; c: Cardinality @ \\
\t1		properties(repetition(x,c)) = properties(x)
\end{axdef}
\begin{itemize}
\item An empty shape expression has no properties.
\item The properties of a triple constraint shape expression are the properties of its triple constraint.
\item Inverse triple constraint shape expressions have no properties.
\item The properties of a some-of, one-of, or grouping shape expression are the union of the properties of their
component shape expressions.
\item The properties of a repetition shape expression are the properties of the shape expression being repeated.
\end{itemize}

Given a triple constraint $tc$, let $propertiesTripleConstraint(tc)$ be its set of properties.
\begin{axdef}
	propertiesTripleConstraint: TripleConstraint \fun PropertiesSet
\where
	\forall a: IRI; C: Constraint @ \\
\t1		propertiesTripleConstraint((nop(a),C)) = \{ a \}
\end{axdef}
\begin{itemize}
\item The properties of a triple constraint is the singleton set that contains its IRI.
\end{itemize}

\subsection{$invproperties$}

Given a shape expression $x$, let $invproperties(x)$ be the set of properties that appear in some 
inverse triple constraint in $x$.
\begin{axdef}
	invproperties: ShapeExpr \fun PropertiesSet
\where
	invproperties(emptyshape) = \emptyset
\also
	\forall tc: TripleConstraint; c: Cardinality @ \\
\t1		invproperties(triple(tc,c)) = \\
\t2			\emptyset
\also
	\forall itc: InverseTripleConstraint; c: Cardinality @ \\
\t1		invproperties(triple(itc,c)) = \\
\t2			invpropertiesInverseTripleConstraint(itc)
\also
	\forall xs: \seq_1 ShapeExpr @ \\
\t1		invproperties(someOf(xs)) = \\
\t1		invproperties(oneOf(xs)) = \\
\t1		invproperties(group(xs)) = \\
\t2			\bigcup \{~ x: \ran xs @ invproperties(x) ~\}
\also
	\forall x: ShapeExpr; c: Cardinality @ \\
\t1		invproperties(repetition(x,c)) = invproperties(x)
\end{axdef}
\begin{itemize}
\item An empty shape expression has no inverse properties.
\item A triple constraint shape expression has no inverse properties.
\item The inverse properties of an inverse triple constraint shape expression are the inverse properties in its inverse triple constraint.
\item The inverse properties of a some-of, one-of, or grouping shape expression is the union of the inverse properties of their
component shape expressions.
\item The inverse properties of a repetition shape expression are the inverse properties of the shape expression being repeated.
\end{itemize}

Given an inverse triple constraint $itc$, let $invpropertiesInverseTripleConstraint(tc)$ be its set of inverse properties.
\begin{axdef}
	invpropertiesInverseTripleConstraint: \\
\t1		InverseTripleConstraint \fun PropertiesSet
\where
	\forall a: IRI; C: ShapeConstr @ \\
\t1		invpropertiesInverseTripleConstraint((inv(a),C)) = \{ a \}
\end{axdef}
\begin{itemize}
\item The inverse properties of an inverse triple constraint is the singleton set that contains its IRI.
\end{itemize}

\subsection{$dep\_graph$}

\subsubsection{$DiGraph$}
A directed graph consists of a set of nodes and a set of directed edges that connect the nodes.
\begin{schema}{DiGraph}[X]
	nodes: \power X \\
	edges : X \rel X
\where
	edges \in nodes \rel nodes
\end{schema}
\begin{itemize}
\item Each edge connects a pair of nodes in the graph.
\end{itemize}

\subsubsection{$DepGraph$}
Given a well-formed schema $S$, let the shapes dependency graph be the directed graph 
whose nodes are the shape labels in $S$ and whose edges connect label $T1$ to label $T2$ when the shape expression 
that defines $T1$ refers to $T2$.\begin{schema}{DepGraph}
	S: SchemaWF \\
	DiGraph[ShapeLabel]
\where
	nodes = shapes(S)
\also
	edges =  \{~ T1, T2 : nodes | T2 \in refsShapeExpr(expr(T1,S)) ~\}
\end{schema}
\begin{itemize}
\item The nodes are the shapes of the schema.
\item There is an edge from $T1$ to $T2$ when the definition of $T1$ refers to $T2$.
\end{itemize}

\subsubsection{$dep\_graph$}
Let $dep\_graph(S)$ be the dependency graph of $S$.
\begin{axdef}
	dep\_graph: SchemaWF \fun DiGraph[ShapeLabel]
\where
	dep\_graph = \{~ DepGraph @ S \mapsto \theta DiGraph ~\}
\end{axdef}

\subsection{$dep\_subgraph$}

\subsubsection{$reachable$}
Given a directed graph $g$ and a node $T$ in $g$, a node $U$ is reachable from $T$ if there is a directed path of one or more edges that connects
$T$ to $U$.
\begin{gendef}[X]
	reachable : DiGraph[X] \cross X \fun \power X
\where
	\forall g: DiGraph[X]; T: X @ \\
\t1		\LET edges == g.edges @ \\
\t2			reachable(g, T) = \{~ U: X | T \mapsto U \in edges\plus ~\}
\end{gendef}

\subsubsection{$DepSubgraph$}
Given a well-formed schema $S$ and a shape label $T$ in $S$, the shapes dependency
graph is the subgraph induced by the nodes that are reachable from $T$.
\begin{schema}{DepSubgraph}
	S: SchemaWF \\
	T: ShapeLabel \\
	DiGraph[ShapeLabel]
\where
	T \in shapes(S)
\also
	\LET g == dep\_graph(S) @ \\
\t1		nodes = reachable(g,T) \land \\
\t1		edges = g.edges \cap (nodes \cross nodes)
\end{schema}
\begin{itemize}
\item The nodes of the subgraph consist of all the nodes reachable from $T$.
\item The edges of the subgraph consist of all edges of the graph whose nodes are in the subgraph.
\end{itemize}

Note that the above formal definition of the dependency subgraph is a literal translation of the text in the semantics draft.
In particular, this literal translation does not explicitly include the label $T$ as a node.
Therefore $T$ will not be in the subgraph unless it is in a directed cycle of edges.

\subsubsection{$dep\_subgraph$}
Let $dep\_subgraph(T,S)$ be the dependency subgraph of $T$ in $S$.
\begin{axdef}
	dep\_subgraph : ShapeLabel \cross SchemaWF \pfun DiGraph[ShapeLabel]
\where
	dep\_subgraph = \{~ DepSubgraph @ (T,S) \mapsto \theta DiGraph ~\}
\end{axdef}

\subsection{$negshapes$}
The definition of $negshapes$ makes use of several auxilliary definitions. 
In the following we assume that $S$ is a well-formed schema and that $T$ is a shape label in $S$.

\subsubsection{$inNeg$}
Let $inNeg(S)$ be the set of labels that appear in some negated shape constraint.
\begin{axdef}
	inNeg: SchemaWF \fun \finset ShapeLabel
\where
	\forall S: SchemaWF @ \\
\t1		inNeg(S) = \bigcup \{~ T: shapes(S) @ inNegExpr(expr(T,S)) ~\}
\end{axdef}

Given a shape expression $x$, let $inNegExpr(x)$ be the set of labels that appear in some negated shape constraint in $x$.
\begin{axdef}
	inNegExpr: ShapeExpr \fun \finset ShapeLabel
\where
	inNegExpr(emptyshape) = \emptyset
\also
	\forall tc: TripleConstraint; c: Cardinality @ \\
\t1		inNegExpr(triple(tc,c)) = \\
\t2			inNegTripleConstraint(tc)
\also
	\forall itc: InverseTripleConstraint; c: Cardinality @ \\
\t1		inNegExpr(triple(itc,c)) = \\
\t2			inNegInverseTripleConstraint(itc)
\also
	\forall xs: \seq_1 ShapeExpr @ \\
\t1		inNegExpr(someOf(xs)) = \\
\t1		inNegExpr(oneOf(xs)) = \\
\t1		inNegExpr(group(xs)) = \\
\t2			\bigcup \{~ x: \ran xs @ inNegExpr(x) ~\}
\also
	\forall x: ShapeExpr; c: Cardinality @ \\
\t1		inNegExpr(repetition(x,c)) = inNegExpr(x)
\end{axdef}

Given a triple constraint $tc$, let $inNegTripleConstraint(tc)$ be the set of labels that appear in some negated shape constraint in $tc$.
\begin{axdef}
	inNegTripleConstraint: TripleConstraint \fun \finset ShapeLabel
\where
	\forall a: IRI; C: ValueConstr @ \\
\t1		inNegTripleConstraint((nop(a),C)) = \emptyset
\also
	\forall a: IRI; C: ShapeConstr @ \\
\t1		inNegTripleConstraint((nop(a),C)) = inNegShapeConstr(C)
\end{axdef}

Given an inverse triple constraint $itc$, let $inNegInverseTripleConstraint(tc)$ be the set of labels that appear in some negated shape constraint in $itc$.
\begin{axdef}
	inNegInverseTripleConstraint: \\
\t1		InverseTripleConstraint \fun \finset ShapeLabel
\where
	\forall a: IRI; C: ShapeConstr @ \\
\t1		inNegInverseTripleConstraint((inv(a),C)) = inNegShapeConstr(C)
\end{axdef}

Given a shape constraint $C$, let $inNegShapeConstr(C)$ be the set of labels that appear in $C$ when it is negated, or the
empty set otherwise.
\begin{axdef}
	inNegShapeConstr: ShapeConstr \fun \finset ShapeLabel
\where
	\forall Ts: \seq_1 ShapeLabel @ \\
\t1		inNegShapeConstr(or(Ts)) = \\
\t1		inNegShapeConstr(and(Ts)) = \\
\t2			\emptyset
\also
	\forall Ts: \seq_1 ShapeLabel @ \\
\t1		inNegShapeConstr(nor(Ts)) = \\
\t1		inNegShapeConstr(nand(Ts)) = \\
\t2			\ran Ts
\end{axdef}

\subsubsection{$underOneOf$}
Let $underOneOf(S)$ be the set of labels that appear in some triple constraint or inverse triple constraint under a one-of constraint
in $S$.
\begin{axdef}
	underOneOf: SchemaWF \fun \finset ShapeLabel
\where
	\forall S: SchemaWF @ \\
\t1		underOneOf(S) = \\
\t2		\bigcup \{~ T: shapes(S) @ underOneOfExpr(expr(T,S)) ~\}
\end{axdef}

Given a shape expression $x$, let $underOneOfExpr(x)$ be the set of labels that appear in some triple constraint or inverse triple constraint under a one-of constraint in $x$.
\begin{axdef}
	underOneOfExpr: ShapeExpr \fun \finset ShapeLabel
\where
	\forall x: ShapeExpr @ \\
\t1		underOneOfExpr(x) = \\
\t2			\IF x \in \ran someOf \\
\t3				\THEN refsShapeExpr(x) \\
\t3				\ELSE \emptyset
\end{axdef}

\subsubsection{$inTripleConstr$}
Let $inTripleConstr(S)$ be the set of labels $T$ such that there is a shape label $T1$ and a triple constraint 
{\tt p::C} or an inverse shape triple constraint
{\tt \verb+^+p::C} in $expr(T1, S)$, and $T$ appears in $C$.

Note that this definition looks wrong since it does not involve negation of shapes.
Nevertheless, a literal translation is given here.
The only difference between $inTripleConstr(S)$ and $refs(S)$ seems to be that the cardinality on the triple and inverse triple
constraints is {\tt [1,1]} since it is not explicitly included in the notations {\tt p::C} and {\tt \verb+^+p::C}.
\begin{axdef}
	inTripleConstr: SchemaWF \fun \finset ShapeLabel
\where
	\forall S: SchemaWF @ \\
\t1		inTripleConstr(S) = \\
\t2			\bigcup \{~ T1: shapes(S) @ inTripleConstrExpr(expr(T1,S)) ~\}
\end{axdef}

Given a shape expression $x$, let $inTripleConstrExpr(x)$ be the set of labels $T$ such that $x$ contains a triple constraint 
{\tt p::C} or an inverse shape triple constraint {\tt \verb+^+p::C} and $T$ appears in $x$.
\begin{axdef}
	inTripleConstrExpr: ShapeExpr \fun \finset ShapeLabel
\where
	inTripleConstrExpr(emptyshape) = \emptyset
\also
	\forall dtc: DirectedTripleConstraint; c: Cardinality @ \\
\t1		inTripleConstrExpr(triple(dtc,c)) = \\
\t2			\IF c = one \\
\t3				\THEN refsDirectedTripleConstraint(dtc) \\
\t3				\ELSE \emptyset
\also
	\forall xs: \seq_1 ShapeExpr @ \\
\t1		inTripleConstrExpr(someOf(xs)) = \\
\t1		inTripleConstrExpr(oneOf(xs)) = \\
\t1		inTripleConstrExpr(group(xs)) = \\
\t2			\bigcup \{~ x: \ran xs @ inTripleConstrExpr(x) ~\}
\also
	\forall x: ShapeExpr; c: Cardinality @ \\
\t1		inTripleConstrExpr(repetition(x,c)) = inTripleConstrExpr(x)
\end{axdef}

\subsubsection{$negshapes$}
The semantics draft makes the following statement.
\begin{quote}
Intuitively, negshapes(S) is the set of shapes labels for which one needs to check whether some nodes in a graph do not satisfy these shapes, in order to validate the graph against the schema S.
\end{quote}

Let $negshapes(S)$ be the set of negated shape labels that appear in $S$.
\begin{axdef}
	negshapes: SchemaWF \fun \finset ShapeLabel
\where
	\forall S: SchemaWF @ \\
\t1		negshapes(S) = inNeg(S) \cup underOneOf(S) \cup inTripleConstr(S)
\end{axdef}
\begin{itemize}
\item A negated shape label is a shape label that appears in a negated shape constraint, or in a triple or inverse triple constraint under a one-of shape expression, or in a triple or inverse triple constraint that has cardinality {\tt [1,1]}.
\end{itemize}
Note that, as remarked above, the definition of $inTripleConstr$ seems wrong.

\subsection{$ShapeVerdict$}
The semantics draft defines the notation {\tt !T} for shape labels $T$ to indicate that $T$ is negated.
The semantics of a schema involves assigning sets of shape labels and negated shape labels to the nodes of a graph, 
which indicates which shapes must be satisfied or violated at each node. 

A shape verdict indicates if a shape must be satisfied or violated.
An asserted label must be satisfied.
A negated label must be violated.
\begin{zed}
	ShapeVerdict ::= \\
\t1		assert\ldata ShapeLabel \rdata | \\
\t1		negate\ldata ShapeLabel \rdata
\end{zed}

The notation {\tt !T} corresponds to $negate(T)$.

\subsection{$allowed$}
Given a value constraint $V$, let $allowed(V)$ be the set of all allowed values defined by $V$.
\begin{axdef}
	allowed: ValueConstr \fun \power (Lit \cup IRI)
\where
	\forall vs: \power (Lit \cup IRI) @ \\
\t1		allowed(valueSet(vs)) = vs
\also
	\forall dt: LiteralDatatype @ \\
\t1		allowed(datatype(dt,\emptyset)) = literalsOfDatatype(dt)
\also
	\forall dt:LiteralDatatype; f: XSFacet @ \\
\t1		allowed(datatype(dt,\{f\})) = literalsOfFacet(dt,f)
\also
	\forall k: NodeKind @ \\
\t1		allowed(kind(k)) = termsOfKind(k)
\end{axdef}

\subsubsection{DAG}
A directed, acyclic graph is a directed graph in which no node is reachable from itself.
\begin{schema}{DAG}[X]
	 DiGraph[X]
\where
	\LET g == \theta DiGraph @ \\
\t1		\forall T: nodes @ T \notin reachable(g,T)
\end{schema}

\subsection{$ReplaceShape$}
The semantics draft introduces the notation $S_{ri}$ for a reduced schema where $S$ is a schema, 
$r$ is a rule-of-one node in a proof tree, and $i$ corresponds to a premise of $r$.
The reduced schema is constructed by replacing a shape with one in which the corresponding one-of component is eliminated.
This replacement operation is described here.
The full definition of $S_{ri}$ is given below following the definition of proof trees.

Given a schema $S$, a shape label $T$ defined in $S$, and a shape expression $Expr'$, the schema $replaceShape(S,T,Expr')$
is the schema $S'$ that is the same as $S$ except that $expr(T,S') = Expr'$.
\begin{schema}{ReplaceShape}
	S, S': SchemaWD \\
	T: ShapeLabel \\
	Expr': ShapeExpr \\
	l: \nat_1\\
	d, d': ShapeDefinition \\
	ecs: \seq ExtensionCondition
\where
	l \in \dom S
\also
	S(l) = (T, d, ecs)
\also
	\forall o: OPTIONAL[InclPropSet]; Expr: ShapeExpr | \\
\t1		d = open(o, Expr) @ \\
\t2			d' = open(o, Expr')
\also
	\forall Expr: ShapeExpr | \\
\t1		d = close(Expr) @ \\
\t2			d' = close(Expr')
\also
	S' = S \oplus \{ l \mapsto (T, d', ecs) \}
\end{schema}

\begin{axdef}
	replaceShape: \\
\t1		SchemaWF \cross ShapeLabel \cross ShapeExpr \pfun SchemaWF
\where
	replaceShape = \{~ ReplaceShape@ (S, T, Expr') \mapsto S' ~\}
\end{axdef}

\subsection{$SchemaWD$}
Given a well-formed schema $S$, it is said to be well-defined if for each negated label $T$ in $negshapes(T)$, the dependency subgraph
$dep\_subgraph(T,S)$ is a directed, acyclic graph.
\begin{zed}
	SchemaWD == \\
\t1		\{~ S: SchemaWF | \\
\t2			\forall T: negshapes(S) @ \\
\t3				dep\_subgraph(T,S) \in DAG[ShapeLabel] ~\}
\end{zed}

The semantics of shape expression schemas is sound only for well-defined schemas.
Only well-defined schemas will be considered from this point forward.

\section{Declarative semantics of shape expression schemas}
\label{sec-declarative-semantics}

Recall that negated triple and inverse triple shape expressions are represented by the corresponding non-negated expressions with cardinality
$none =$ {\tt [0;0]}.

\subsection{$LabelledTriple$}
A labelled triple is either an incoming or outgoing edge in an RDF graph.
\begin{zed}
	LabelledTriple ::= \\
\t1		out \ldata Triple \rdata | \\
\t1		inc \ldata Triple \rdata
\end{zed}

Sometimes labelled triples are referred to simply as triples.

\subsection{$matches$}
A labelled triple matches a directed triple constraint when they have the same direction and predicate.
\begin{axdef}
	matches: LabelledTriple \rel DirectedTripleConstraint
\where
	matches = matches\_out \cup matches\_inc
\end{axdef}

\subsubsection{$matches\_out$}
$matches\_out$ matches outgoing triples to triple constraints.
\begin{zed}
	matches\_out == \\
\t1		\{~ s, p, o: TERM; C: Constraint | \\
\t2			(s,p,o) \in Triple @ \\
\t3				out(s,p,o) \mapsto (nop(p),C) ~\}
\end{zed}

Note that this definition ignores any value constraints defined in $C$.
The absence of restrictions imposed by value constraints makes matching weaker than it could be.
This may be an error in the semantics draft.

The semantics drafts contains the following text.
\begin{quote}
The following definition introduces the notion of satisfiability of a shape constraint by a set of triples. Such satisfiability is going to be used for checking that the neighbourhood of a node satisfies locally the constraints defined by a shape expression, without taking into account whether the shapes required by the triple constraints and inverse triple constraints are satisfied.
\end{quote}

Read literally, only shape constraints should be ignored, so unless value constraints are handled elsewhere, 
the semantics draft has an error in the definition of $matches$.

\subsubsection{$matches\_inc$}
$matches\_inc$ matches incoming triples to inverse triple constraints.
\begin{zed}
	matches\_inc == \\
\t1		\{~ s, p, o: TERM; C: ShapeConstr | \\
\t2			(s,p,o) \in Triple @ \\
\t3				inc(s,p,o) \mapsto (inv(p),C) ~\} 
\end{zed}

\subsection{$satifies$}
A set of labelled triples $Neigh$ is said to satisfy a shape expression $Expr$ if the constraints, other than shape constraints,
defined in $Expr$ are satisfied.

Note that the definition of $matches$ ignores both value and shape constraints.

\begin{axdef}
	satisfies: \finset LabelledTriple \rel ShapeExpr
\end{axdef}

This relation is defined recursively by inference rules for each type of shape expression.
\begin{zed}
	satisfies = \\
\t1		rule\_empty \cup \\
\t1		rule\_triple\_constraint \cup \\
\t1		rule\_inverse\_triple\_constraint \cup \\
\t1		rule\_some\_of \cup \\
\t1		rule\_one\_of \cup \\
\t1		rule\_group \cup \\
\t1		rule\_repeat
\end{zed}

\subsubsection{$InfRule$}
An inference rule defines a relation between a set of labelled triples and a shape expression.
It is convenient to define a base schema for the inference rules.
\begin{schema}{InfRule}
	Neigh: \finset LabelledTriple \\
	Expr: ShapeExpr
\end{schema}

\subsubsection{$rule\_empty$}
An empty set of triples satisfies the empty shape expression.
\begin{schema}{RuleEmpty}
	InfRule
\where
	Expr = emptyshape
\also
	Neigh = \emptyset
\end{schema}

\begin{axdef}
	rule\_empty: \finset LabelledTriple \rel ShapeExpr
\where
	rule\_empty = \\ 
\t1		\{~ RuleEmpty @ Neigh \mapsto Expr ~\}
\end{axdef}

\subsubsection{$rule\_triple\_constraint$}
A set of triples satisfies a triple constraint shape expression when each triple matches the constraint
and the total number of constraints is within the bounds of the cardinality.
\begin{schema}{RuleTripleConstraint}
	InfRule \\
	k: \nat \\
	p: IRI \\
	C: Constraint \\
	c: Cardinality
\where
	Expr = triple((nop(p),C),c)
\also
	k = \# Neigh
\also
	k \inrel{inBounds} c
\also
	\forall t: Neigh @ t \inrel{matches} (nop(p),C)
\end{schema}

\begin{axdef}
	rule\_triple\_constraint: \finset LabelledTriple \rel ShapeExpr
\where
	rule\_triple\_constraint = \\
\t1		\{~ RuleTripleConstraint @ Neigh \mapsto Expr ~\}
\end{axdef}

\subsubsection{$rule\_inverse\_triple\_constraint$}
A set of triples satisfies an inverse triple constraint shape expression when each triple matches the constraint
and the total number of constraints is within the bounds of the cardinality.

\begin{schema}{RuleInverseTripleConstraint}
	InfRule \\
	k: \nat \\
	p: IRI \\
	C: Constraint \\
	c: Cardinality
\where
	Expr = triple((inv(p),C),c)
\also
	k = \# Neigh
\also
	k \inrel{inBounds} c
\also
	\forall t: Neigh @ t \inrel{matches} (inv(p),C)
\end{schema}

\begin{axdef}
	rule\_inverse\_triple\_constraint: \finset LabelledTriple \rel ShapeExpr
\where
	rule\_triple\_constraint = \\
\t1		\{~ RuleInverseTripleConstraint @ Neigh \mapsto Expr ~\}
\end{axdef}

\subsubsection{$rule\_some\_of$}
A set of triples satisfies a some-of shape expression when the set of triples satisfies one of the component
shape expressions.
\begin{schema}{RuleSomeOf}
	InfRule \\
	Exprs: \seq_1 ShapeExpr \\
	i: \nat
\where
	Expr = someOf(Exprs)
\also
	i \in \dom Exprs
\also
	Neigh \inrel{satisfies} Exprs(i)
\end{schema}

\begin{axdef}
	rule\_some\_of: \finset LabelledTriple \rel ShapeExpr
\where
	rule\_some\_of = \\
\t1		\{~ RuleSomeOf @ Neigh \mapsto Expr ~\}
\end{axdef}

\subsubsection{$rule\_one\_of$}
A set of triples satisfies a one-of shape expression when the set of triples satisfies one of the component
shape expressions.
\begin{schema}{RuleOneOf}
	InfRule \\
	Exprs: \seq_1 ShapeExpr \\
	i: \nat
\where
	Expr = oneOf(Exprs)
\also
	i \in \dom Exprs
\also
	Neigh \inrel{satisfies} Exprs(i)
\end{schema}

\begin{axdef}
	rule\_one\_of: \finset LabelledTriple \rel ShapeExpr
\where
	rule\_one\_of = \\
\t1		\{~ RuleOneOf @ Neigh \mapsto Expr ~\}
\end{axdef}

The semantics draft contains the following text.
\begin{quote}
Note that the conditions for some-of and one-of shapes are identical. 
The distinction between both will be made by taking into account also the non-local, shape constraints.
\end{quote}

\subsubsection{$rule\_group$}
A set of triples satisfies a group shape expression when the set of triples can be partitioned into a sequence of subsets whose length
is the same as the sequence of component shape expressions, and each subset satisfies the corresponding component shape expression.
\begin{schema}{RuleGroup}
	InfRule \\
	Neighs: \seq_1 (\finset LabelledTriple) \\
	Exprs: \seq_1 ShapeExpr
\where
	Expr = group(Exprs)
\also
	Neighs \partition Neigh
\also
	\# Neighs = \# Exprs
\also
	\forall j: \dom Neighs @ \\
\t1		Neighs(j) \inrel{satisfies} Exprs(j)
\end{schema}

\begin{axdef}
	rule\_group: \finset LabelledTriple \rel ShapeExpr
\where
	rule\_group = \\
\t1		\{~ RuleGroup @ Neigh \mapsto Expr ~\}
\end{axdef}

\subsubsection{$rule\_repeat$}
A set of triples satisfies a repetition shape expression when the set of triples can be partitioned into a sequence of subsets whose length is
in the bounds of the cardinality, and each subset satisfies the component shape expression of the repetition shape expression.
\begin{schema}{RuleRepeat}
	InfRule \\
	Expr1: ShapeExpr \\
	Neighs: \seq_1 (\finset LabelledTriple) \\
	k: \nat \\
	c: Cardinality
\where
	Expr = repetition(Expr1,c)
\also
	k = \# Neighs
\also
	k \inrel{inBounds} c
\also
	Neighs \partition Neigh
\also
	\forall j: \dom Neighs @ \\
\t1		Neighs(j) \inrel{satisfies} Expr1
\end{schema}

\begin{axdef}
	rule\_repeat: \finset LabelledTriple \rel ShapeExpr
\where
	rule\_repeat = \\
\t1		\{~ RuleRepeat @ Neigh \mapsto Expr ~\}
\end{axdef}

\subsection{Proof Trees}
The preceding definition of $satisfies$ is based on the existence of certain characteristics of the set of triples.
For example, a set of triples satisfies one of a sequence of shape expressions when it satisfies exactly one of the them, 
but the $satisfies$ relation forgets the actual shape expression that the set of triples satisfies.
We can remember this type of information in a proof tree.

\subsubsection{$RuleTree$}
A rule tree is a tree of inference rules and optional child rule trees.
Child rule trees occur in cases where the inference rule depends on other inference rules.
\begin{zed}
	RuleTree ::= \\
\t1		ruleEmpty \ldata RuleEmpty \rdata | \\
\t1		ruleTripleConstraint \ldata RuleTripleConstraint \rdata | \\
\t1		ruleInverseTripleConstraint \ldata RuleInverseTripleConstraint \rdata | \\
\t1		ruleSomeOf \ldata RuleSomeOf \cross RuleTree \rdata | \\
\t1		ruleOneOf \ldata RuleOneOf \cross RuleTree \rdata | \\
\t1		ruleGroup \ldata RuleGroup \cross \seq_1 RuleTree \rdata | \\
\t1		ruleRepeat \ldata RuleRepeat \cross \seq_1 RuleTree \rdata
\end{zed}

\subsubsection{$baseRule$}
Each node in a rule tree contains an inference rule and, therefore, a base inference rule.
\begin{axdef}
	baseRule: RuleTree \fun InfRule
\where
	\forall RuleEmpty @ \\
\t1		\LET rule == \theta RuleEmpty; \\
\t2			base == \theta InfRule @ \\
\t3				baseRule(ruleEmpty(rule)) = base
\also
	\forall RuleTripleConstraint @ \\
\t1		\LET rule == \theta RuleTripleConstraint; \\
\t2			base == \theta InfRule @ \\
\t3			baseRule(ruleTripleConstraint(rule)) = base
\also
	\forall RuleInverseTripleConstraint @ \\
\t1		\LET rule == \theta RuleInverseTripleConstraint; \\
\t2			base == \theta InfRule @ \\
\t3				baseRule(ruleInverseTripleConstraint(rule)) = base
\also
	\forall RuleSomeOf; tree: RuleTree @ \\
\t1		\LET rule == \theta RuleSomeOf; \\
\t2			base == \theta InfRule @ \\
\t3				baseRule(ruleSomeOf(rule,tree)) = base
\also
	\forall RuleOneOf; tree: RuleTree @ \\
\t1		\LET rule == \theta RuleOneOf; \\
\t2			base == \theta InfRule @ \\
\t3				baseRule(ruleOneOf(rule,tree)) = base
\also
	\forall RuleGroup; trees: \seq_1 RuleTree @ \\
\t1		\LET rule == \theta RuleGroup; \\
\t2			base == \theta InfRule @ \\
\t3				baseRule(ruleGroup(rule,trees)) = base
\also
	\forall RuleRepeat; trees: \seq_1 RuleTree @ \\
\t1		\LET rule == \theta RuleRepeat; \\
\t2			base == \theta InfRule @ \\
\t3				baseRule(ruleRepeat(rule,trees)) = base
\end{axdef}

\subsubsection{$baseNeigh$}
Each node in a rule tree has a base set of labelled triples.
\begin{axdef}
	baseNeigh: RuleTree \fun \finset LabelledTriple
\where
	\forall tree: RuleTree @ \\
\t1		baseNeigh(tree) = (baseRule(tree)).Neigh
\end{axdef}

\subsubsection{$baseExpr$}
Each node in a rule tree has a base shape expression.
\begin{axdef}
	baseExpr: RuleTree \fun ShapeExpr
\where
	\forall tree: RuleTree @ \\
\t1		baseExpr(tree) = (baseRule(tree)).Expr
\end{axdef}

\subsubsection{$ProofTree$}
A proof tree is a rule tree in which the child trees prove subgoals of their parent nodes.
\begin{axdef}
	ProofTree: \power RuleTree
\end{axdef}

The definition of proof tree is recursive so it is given by a set of constraints, one for each type of node.

Any rule tree whose root node contains an empty shape expression is a proof tree
since it has no subgoals.
\begin{zed}
	\ran ruleEmpty \subset ProofTree
\end{zed}

Any rule tree whose root node node contains a triple constraint shape expression is a proof tree
since it has no subgoals.
\begin{zed}
	\ran ruleTripleConstraint \subset ProofTree
\end{zed}

Any rule tree whose root node node contains an inverse triple constraint shape expression is a proof tree
since it has no subgoals.
\begin{zed}
	\ran ruleInverseTripleConstraint \subset ProofTree
\end{zed}

A rule tree whose root node contains a some-of shape expression is a proof tree if and only if
its child rule tree correspond to the distinguished shape expression at index $i$ and it is a proof tree.
\begin{zed}
	\forall RuleSomeOf; tree: RuleTree @ \\
\t1		ruleSomeOf(\theta RuleSomeOf, tree) \in ProofTree \iff \\
\t2			baseNeigh(tree) = Neigh \land \\
\t2			baseExpr(tree) = Exprs(i) \land \\
\t2			tree \in ProofTree
\end{zed}

A rule tree whose root node contains a one-of shape expression is a proof tree if and only if
its child rule tree correspond to the distinguished shape expression at index $i$ and it is a proof tree.
\begin{zed}
	\forall RuleOneOf; tree: RuleTree @ \\
\t1		ruleOneOf(\theta RuleOneOf, tree) \in ProofTree \iff \\
\t2			baseNeigh(tree) = Neigh \land \\
\t2			baseExpr(tree) = Exprs(i) \land \\
\t2			tree \in ProofTree
\end{zed}

A rule tree whose root node contains a group shape expression is a proof tree if and only if
its sequence of child rule trees correspond to its sequence of component neighbourhood and shape expressions
and each child rule tree is a proof tree.
\begin{zed}
	\forall RuleGroup; trees: \seq_1 RuleTree @ \\
\t1		ruleGroup(\theta RuleGroup, trees) \in ProofTree \iff \\
\t2			\# Exprs = \# trees \land \\
\t2			(\forall i: \dom trees @ \\
\t3				baseNeigh(trees(i)) = Neighs(i) \land \\
\t3				baseExpr(trees(i)) = Exprs(i) \land \\
\t3				trees(i) \in ProofTree)
\end{zed}

A rule tree whose root node contains a repetition shape expression is a proof tree if and only if
its sequence of child rule trees correspond to its sequence of component neighbourhoods
and each child rule tree is a proof tree.
\begin{zed}
	\forall RuleRepeat; trees: \seq_1 RuleTree @ \\
\t1		ruleRepeat(\theta RuleRepeat, trees) \in ProofTree \iff \\
\t2			\# Neighs = \# trees \land \\
\t2			(\forall i: \dom trees @ \\
\t3				baseNeigh(trees(i)) = Neighs(i) \land \\
\t3				baseExpr(trees(i)) = Expr1 \land \\
\t3				trees(i) \in ProofTree)
\end{zed}

We have the following relation between proof trees and the $satisfies$ relation.
\[\vdash
	satisfies = \\
\t1		\{~ tree: ProofTree @ baseNeigh(tree) \mapsto baseExpr(tree) ~\}
\]

\subsection{Reduced Schema for rule-one-of}
As mentioned above, inference rules and proof trees treat rule-one-of exactly the same as rule-some-of.
The difference between these rules appears when considering valid typings, which are described in detail later.

Let $t$ be a valid typing of graph $G$ under schema $S$.
Let $n$ be a node in $G$ and let $T$ be a shape label in $t(n)$.
Let $Expr = expr(T,S)$ be the shape expression for $T$.
Let $tree$ be a proof tree that the neighbourhood of $n$ satisfies $Expr$.
Let $r$ be a node of the proof tree that contains an application of rule-one-of
and let $i$ be the index of the component expression used in the application of the rule.
The intention of the one-of shape expression is that the triples match exactly one of the component expressions.
Therefore, if the matched shape expression is removed from the one-of expression then there must not be any valid typings
of $G$ under the reduced schema $S_{ri}$.

Note that a one-of shape expression may have one or more components.
The number of components is denoted by $k$ in the inference rule.
However, if it contains exactly one component then there no further semantic conditions that must hold and there
is no corresponding reduced schema.
Therefore, the definition of the reduced schema only applies to the case where the number of components is greater than one,
i.e. $k > 1$.

Rule trees are ordered trees.
A child tree can be specified by giving its index among all the children.
The maximum index of a child depends on the type of rule.
For leaf trees, the maximum child index is 0.
\begin{axdef}
	maxChild: RuleTree \fun \nat
\where
	\forall tree: \ran ruleEmpty @ maxChild(tree) = 0
\also
	\forall tree: \ran ruleTripleConstraint @ maxChild(tree) = 0
\also
	\forall tree: \ran ruleInverseTripleConstraint @ maxChild(tree) = 0
\also
	\forall tree: \ran ruleSomeOf @ maxChild(tree) = 1
\also
	\forall tree: \ran ruleOneOf @ maxChild(tree) = 1
\also
	\forall r: RuleGroup; trees: \seq_1 RuleTree @ \\
\t1		maxChild(ruleGroup(r,trees)) = \# trees
\also
	\forall r: RuleRepeat; trees: \seq_1 RuleTree @ \\
\t1		maxChild(ruleRepeat(r,trees)) = \# trees
\end{axdef}

Given a tree $tree$ and a valid child index $j$, the child tree at the index is $childAt(tree,j)$.
\begin{axdef}
	childAt: RuleTree \cross \nat_1 \pfun RuleTree
\where
	\dom childAt = \\
\t1		\{~ tree: RuleTree; ci: \nat_1 | ci \leq maxChild(tree) ~\}
\also
	\forall r: RuleSomeOf; tree: RuleTree @ \\
\t1		childAt(ruleSomeOf(r,tree),1) = tree
\also
	\forall r: RuleOneOf; tree: RuleTree @ \\
\t1		childAt(ruleOneOf(r,tree),1) = tree
\also
	\forall r: RuleGroup; trees: \seq_1 RuleTree @ \\
\t1		\LET tree == ruleGroup(r,trees) @ \\
\t2			\forall ci: 1 \upto maxChild(tree) @ \\
\t3				childAt(tree,ci) = trees(ci)
\also
	\forall r: RuleRepeat; trees: \seq_1 RuleTree @ \\
\t1		\LET tree == ruleRepeat(r,trees) @ \\
\t2			\forall ci: 1 \upto maxChild(tree) @ \\
\t3				childAt(tree,ci) = trees(ci)
\end{axdef}

The location of a node within a rule tree can be specified by giving a sequence of positive integers that specify the index of each child tree.
The root of the tree is specified by the empty sequence.
Such a sequence of integers is referred to as a rule path.
Given a rule tree $tree$, the set of all of its rule paths is $rulePaths(tree)$.
\begin{axdef}
	rulePaths: RuleTree \fun \finset (\seq \nat_1)
\where
	\forall tree: RuleTree | maxChild(tree) = 0 @ \\
\t1		rulePaths(tree) = \{ \langle \rangle \}
\also
	\forall tree: RuleTree | maxChild(tree) > 0 @ \\
\t1		rulePaths(tree) = \\
\t2			\bigcup \{~ ci : 1 \upto maxChild(tree) @ \\
\t3				\{~ path: rulePaths(childAt(tree,ci)) @ \langle ci \rangle \cat path ~\} ~\}
\end{axdef}

Given a rule tree $tree$ and a rule path $path$, the tree node specified by the path is $treeAt(tree,path)$,
\begin{axdef}
	treeAt: RuleTree \cross \seq \nat_1 \pfun RuleTree
\where
	\dom treeAt = \\
\t1		\{~ tree: RuleTree; path: \seq \nat_1 | path \in rulePaths(tree) ~\}
\also
	\forall tree: RuleTree @ treeAt(tree, \langle \rangle) = tree
\also
	\forall tree: RuleTree; ci: \nat_1; path: \seq \nat_1 | \\
\t1		\langle ci \rangle \cat path \in rulePaths(tree) @ \\
\t2			treeAt(tree, \langle ci \rangle \cat path) = treeAt(childAt(tree,ci), path)
\end{axdef}

Given a one-of shape expression $Expr$ that has more than one component, and an index $i$ of one component,
$elimExpr(Expr,i)$ is the reduced expression in which component $i$ is eliminated.
\begin{schema}{ElimExpr}
	Expr, Expr': ShapeExpr \\
	Exprs, ExprsL, ExprsR: \seq_1 ShapeExpr \\
	i: \nat
\where
	Expr = oneOf(Exprs)
\also
	\# Exprs > 1
\also
	i \in \dom Exprs
\also
	Exprs = ExprsL \cat \langle Exprs(i) \rangle \cat ExprsR
\also
	Expr' = oneOf(ExprsL \cat ExprsR)
\end{schema}

\begin{axdef}
	elimExpr: ShapeExpr \cross \nat \pfun ShapeExpr
\where
	elimExpr = \{~ ElimExpr @ (Expr,i) \mapsto Expr' ~\}
\end{axdef}

Given a proof tree $tree$ with the shape expression $Expr$ as its base, and a path $path$ to some application $r$ of rule-one-of in $tree$
in which the rule-of expression has more than one component,
\begin{schema}{RuleOneOfApplication}
	tree: ProofTree \\
	path: \seq \nat_1 \\
	r, rChild: ProofTree \\
	R: RuleOneOf
\where
	path \in rulePaths(tree)
\also
	r = treeAt(tree,path) = ruleOneOf(R,rChild)
\also
	\# R.Exprs > 1
\end{schema}
\begin{itemize}
\item The path is a valid rule path in the proof tree.
\item	The tree at the path is an application of rule-one-of.
\item There are more than one components in the one-of shape expression.
\end{itemize}

$reduceExpr(tree,path)$ is the reduced base shape expression with the corresponding one-of expression in $Expr$ replaced by the reduced one-of expression.
\begin{axdef}
	reduceExpr: ProofTree \cross \seq \nat_1 \pfun ShapeExpr
\where
	\dom reduceExpr = \{~ RuleOneOfApplication @ (tree,path) ~\}
\also
	\forall RuleOneOfApplication | \\
\t1		path = \langle \rangle \land \\
\t1		tree = r @ \\
\t2			reduceExpr(r,\langle \rangle) = elimExpr(R.Expr, R.i)
\end{axdef}
\begin{itemize}
\item The domain of this function requires that the path be a valid rule path in the proof tree.
\item In the case of an empty path, the tree must be a one-of tree and the branch taken is eliminated.
\item When the path is not empty, this function is defined recursively by additional constraints which follow.
There are four possible cases in which the proof tree has children. These cases correspond to applications of
rule-some-of, rule-one-of, rule-group, and rule-repeat.
Each case is defined by a schema below.
\end{itemize}

\begin{schema}{ReduceSomeOf}
	RuleOneOfApplication \\
	RuleSomeOf \\
	child: ProofTree \\
	tail: \seq \nat_1 \\
	ExprsL, ExprsR: \seq ShapeExpr \\
	Expr': ShapeExpr
\where
	tree = ruleSomeOf(\theta RuleSomeOf, child)
\also
	path = \langle 1 \rangle \cat tail
\also
	Exprs = ExprsL \cat \langle Exprs(i) \rangle \cat ExprsR
\also
	Expr' = someOf(ExprsL \cat \langle reduceExpr(child,tail) \rangle \cat ExprsL)
\end{schema}

\begin{zed}
	\forall ReduceSomeOf @ \\
\t1		reduceExpr(tree,path) = Expr'
\end{zed}

\begin{schema}{ReduceOneOf}
	RuleOneOfApplication \\
	RuleOneOf \\
	child: ProofTree \\
	tail: \seq \nat_1 \\
	ExprsL, ExprsR: \seq ShapeExpr \\
	Expr': ShapeExpr
\where
	tree = ruleOneOf(\theta RuleOneOf, child)
\also
	path = \langle 1 \rangle \cat tail
\also
	Exprs = ExprsL \cat \langle Exprs(i) \rangle \cat ExprsR
\also
	Expr' = oneOf(ExprsL \cat \langle reduceExpr(child,tail) \rangle \cat ExprsL)
\end{schema}

\begin{zed}
	\forall ReduceOneOf @ \\
\t1		reduceExpr(tree,path) = Expr'
\end{zed}

\begin{schema}{ReduceGroup}
	RuleOneOfApplication \\
	RuleGroup \\
	children: \seq_1 ProofTree \\
	ci: \nat_1 \\
	tail: \seq \nat_1 \\
	ExprsL, ExprsR: \seq ShapeExpr \\
	Expr': ShapeExpr
\where
	tree = ruleGroup(\theta RuleGroup, children)
\also
	path = \langle ci \rangle \cat tail
\also
	Exprs = ExprsL \cat \langle Exprs(ci) \rangle \cat ExprsR
\also
	Expr' = group(ExprsL \cat \langle reduceExpr(children(ci),tail) \rangle \cat ExprsL)
\end{schema}

\begin{zed}
	\forall ReduceGroup @ \\
\t1		reduceExpr(tree,path) = Expr'
\end{zed}

\begin{schema}{ReduceRepeat}
	RuleOneOfApplication \\
	RuleRepeat \\
	children: \seq_1 ProofTree \\
	ci: \nat_1 \\
	tail: \seq \nat_1 \\
	Expr': ShapeExpr
\where
	tree = ruleRepeat(\theta RuleRepeat, children)
\also
	path = \langle ci \rangle \cat tail
\also
	Expr' = repetition(reduceExpr(children(ci),tail),c)
\end{schema}

\begin{zed}
	\forall ReduceRepeat @ \\
\t1		reduceExpr(tree,path) = Expr'
\end{zed}
\begin{itemize}
\item Something looks wrong here because if a repetition expression has a one-of expression as a child then there is no way
to associate the reduced one-of expression with just the path taken in the proof tree since all the children of a repetition expression
share the same shape expression. However, a rule-repeat node in the proof tree has many children and there is no requirement
that all children would use the same branch of the one-of expression.
To make progress, I'll assume that all children of the repeat will eliminate the same branch of the one-of.
I will report this to the mailing list later, along with the observation that the reduction should only one done when a one-of expression
has more than one component.
\end{itemize}

\subsection{Witness Mappings}
Given a set of labelled triples $Neigh$, a shape expression $Expr$ and a proof tree $tree$ that proves $Neigh$ satisfies $Expr$, 
each labelled triple $triple$ appears in a unique leaf node of the proof tree whose rule matches $triple$ with a directed triple constraint $dtc$.
This association of $triple$ with $dtc$ is called a witness mapping, $wm(triple) = dtc$.

\subsection{$WitnessMapping$}
\begin{zed}
	WitnessMapping == LabelledTriple \pfun DirectedTripleConstraint
\end{zed}

\subsubsection{$witness$}
\begin{axdef}
	witness: ProofTree \fun WitnessMapping
\where
	\forall r: RuleEmpty @ \\
\t1		\LET tree == ruleEmpty(r) @ \\
\t2		witness(tree) = \emptyset
\also
	\forall r: RuleTripleConstraint; dtc: DirectedTripleConstraint; c: Cardinality | \\
\t1		r.Expr = triple(dtc,c) @ \\
\t2			\LET tree == ruleTripleConstraint(r) @ \\
\t3				witness(tree) = baseNeigh(tree) \cross \{dtc\}
\also
	\forall r: RuleInverseTripleConstraint; dtc: DirectedTripleConstraint; c: Cardinality | \\
\t1		r.Expr = triple(dtc,c) @ \\
\t2			\LET tree == ruleInverseTripleConstraint(r) @ \\
\t3				witness(tree) = baseNeigh(tree) \cross \{dtc\}
\also
	\forall r: RuleSomeOf; subtree: ProofTree @ \\
\t1		\LET tree == ruleSomeOf(r, subtree) @ \\
\t2			tree \in ProofTree \implies \\
\t3				witness(tree) = witness(subtree)
\also
	\forall r: RuleOneOf; subtree: ProofTree @ \\
\t1		\LET tree == ruleOneOf(r, subtree) @ \\
\t2			tree \in ProofTree \implies \\
\t3				witness(tree) = witness(subtree)
\also
	\forall r: RuleGroup; subtrees: \seq_1 ProofTree @ \\
\t1		\LET tree == ruleGroup(r, subtrees) @ \\
\t2			tree \in ProofTree \implies \\
\t3				witness(tree) = \bigcup \{~ subtree: \ran subtrees @ witness(subtree) ~\}
\also
	\forall r: RuleRepeat; subtrees: \seq_1 ProofTree @ \\
\t1		\LET tree == ruleRepeat(r, subtrees) @ \\
\t2			tree \in ProofTree \implies \\
\t3				witness(tree) = \bigcup \{~ subtree: \ran subtrees @ witness(subtree) ~\}
\end{axdef}

\subsection{$outNeigh$}
The outgoing neighbourhood of a node $n$ in an RDF graph $G$ is the set of outgoing labelled triples that correspond to triples
in $G$ with subject $n$.
\begin{axdef}
	outNeigh: Graph \cross TERM \fun \finset LabelledTriple
\where
	\forall G: Graph; n: TERM @ \\
\t1		outNeigh(G,n) = \{~ p, o: TERM | (n,p,o) \in G @ out(n,p,o) ~\}
\end{axdef}

\subsection{$incNeigh$}
The ingoing neighbourhood of a node $n$ in an RDF graph $G$ is the set of ingoing labelled triples that correspond to triples
in $G$ with object $n$.
\begin{axdef}
	incNeigh: Graph \cross TERM \fun \finset LabelledTriple
\where
	\forall G: Graph; n: TERM @ \\
\t1		incNeigh(G,n) = \{~ p, s: TERM | (s,p,n) \in G @ inc(n,p,s) ~\}
\end{axdef}

\subsection{$Typing$}
Given a schema $S$ and a graph $G$, a typing $t$ is a map that associates to each node $n$ of $G$ a, possibly empty, set $t(n)$ of shape labels and negated shape labels such that if {\tt T} is a negated shape label then either {\tt T} or {\tt !T} is in $t(n)$.
Here I infer that {\tt T} and {\tt !T} are mutually exclusive.

A typing map associates a finite, possibly empty, set of shape verdicts to nodes.
\begin{zed}
	Typing == TERM \pfun \finset ShapeVerdict
\end{zed}

\begin{schema}{TypingMap}
	G: Graph \\
	S: SchemaWD \\
	t: Typing
\where
	\dom t = nodes(G)
\also
	\forall n: nodes(G); T: ShapeLabel | assert(T) \in t(n) @ \\
\t1		T \in shapes(S)
\also
	\forall n: nodes(G); T: ShapeLabel | negate(T) \in t(n) @ \\
\t1		T \in negshapes(S)
\also
	\forall n: nodes(G); T: negshapes(S) @ \\
\t1		assert(T) \in t(n) \lor negate(T) \in t(n)
\also
	\forall n: nodes(G); T: shapes(S) @ \\
\t1		assert(T) \notin t(n) \lor negate(T) \notin t(n)
\end{schema}
\begin{itemize}
\item The typing associates a set of shape verdicts to each node in the graph.
\item If a node is required to satisfy $T$ then $T$ must be a shape label of the schema.
\item If a node is required to violate $T$ then $T$ must be a negated shape label of the schema.
\item If $T$ is a negated shape label of the schema then each node must be required to either satisfy or violate it.
\item No node must be required to both satisfy and violate the same shape.
\end{itemize}

\begin{axdef}
	typings: Graph \cross SchemaWD \fun \power Typing
\where
	\forall G: Graph; S: SchemaWD @ \\
\t1		typings(G,S) = \{~ m: TypingMap | m.G = G \land m.S = S @ m.t ~\}
\end{axdef}

\subsection{$TypingSatisfies$}
Given a typing $t$, a node $u$, and a shape constraint $C$, the typing satisfies the constraint at the node if the 
boolean conditions implied by the shape constraint hold.

\begin{schema}{TypingSatisfies}
	TypingMap \\
	u: TERM \\
	C: ShapeConstr \\
	Ts: \seq_1 ShapeLabel
\where
	u \in nodes(G)
\also
	C = and(Ts) \implies \\
\t1		(\forall T: \ran Ts @ assert(T) \in t(u))
\also
	C = or(Ts) \implies \\
\t1		(\exists T: \ran Ts @ assert(T) \in t(u))
\also
	C = nand(Ts) \implies \\
\t1		(\exists T: \ran Ts @ negate(T) \in t(u))
\also
	C = nor(Ts) \implies \\
\t1		(\forall T: \ran Ts @ negate(T) \in t(u))
\end{schema}
\begin{itemize}
\item The node is in the graph.
\item The node is required to satisfy every shape in an and shape constraint.
\item The node is required to satisfy some shape in an or shape constraint.
\item The node is required to violate some shape in a nand shape constraint.
\item The node is required to violate every shape in a nor shape constraint.
\end{itemize}

\begin{axdef}
	typingSatisfies: Typing \cross TERM \rel ShapeConstr
\where
	typingSatisfies = \\
\t1		\{~ TypingSatisfies @ (t, u) \mapsto C ~\}
\end{axdef}

\subsection{$Matching$}
Given a node $n$ in graph $G$, a typing $t$, and a directed triple constraint $X$, let $Matching(G,n,t,X)$ be the set of triples in the graph with focus node $n$ that match $X$ under $t$.

\begin{schema}{MatchingTriples}
	TypingMap \\
	n, p: TERM \\
	X: DirectedTripleConstraint \\
	C: Constraint \\
	triples: \finset LabelledTriple
\where
	C \in ValueConstr \land X = (nop(p),C) \implies \\
\t1		triples = \{~ u: TERM | (n,p,u) \in G \land \\
\t2			u \in allowed(C)@ out(n,p,u) ~\}
\also
	C \in ShapeConstr \land X = (nop(p),C) \implies \\
\t1		triples = \{~ u: TERM | (n,p,u) \in G \land \\
\t2			(t,u) \inrel{typingSatisfies} C @ out(n,p,u) ~\}
\also
	C \in ShapeConstr \land X = (inv(p),C) \implies \\
\t1		triples = \{~ u: TERM | (u,p,n) \in G \land \\
\t2			(t,u) \inrel{typingSatisfies} C @ inc(u,p,n) ~\}
\end{schema}
\begin{itemize}
\item An outgoing triple matches a value constraint if its object is an allowed value.
\item An outgoing triple matches a shape constraint if the typing of its object satisfies the constraint.
\item An incoming triple matches a shape constraint if the typing of its subject satisfies the constraint.
\end{itemize}

\begin{axdef}
	Matching: Graph \cross TERM \cross Typing \cross DirectedTripleConstraint \fun \\
\t1		\finset LabelledTriple
\where
	Matching = \{~ MatchingTriples @ (G, n, t, X) \mapsto triples ~\}
\end{axdef}

\subsection{$validTypings$}
The definition of what it means for a graph to satisfy a shape schema is given in terms of the existence of a valid typing.
Given a graph $G$ and a schema $S$, a valid typing of $G$ by $S$ is a typing that satisfies certain additional conditions
at each node $n$ in $G$.
\begin{axdef}
	validTypings: Graph \cross SchemaWD \fun \power Typing
\where
	\forall G: Graph; S: SchemaWD @ \\
\t1		validTypings(G,S) \subseteq typings(G,S)
\end{axdef}

\subsubsection{$ValidTypingNodeLabel$}
The definition of a valid typing is given in terms of a series of conditions that must hold at each node and for each shape verdict at that node.
It is convenient to introduce the following base schema for conditions.
\begin{schema}{ValidTypingNodeLabel}
	TypingMap \\
	n: TERM \\
	T: ShapeLabel \\
	ruleT: Rule \\
	defT: ShapeDefinition \\
	Expr: ShapeExpr \\
	Xs: \finset DirectedTripleConstraint
\where
	t \in validTypings(G,S)
\also
	n \in nodes(G)
\also
	assert(T) \in t(n) \lor negate(T) \in t(n)
\also
	ruleT = rule(T,S) \\
	defT = shapeDef(ruleT)
\also
	Expr = expr(T,S) \\
	Xs = tripleConstraints(Expr)
\end{schema}

\subsubsection{$tripleConstraints$}
Given a shape expression $Expr$ let $tripleConstraints(Expr)$ be the set of all triple or inverse triple constraints contained in it.
\begin{axdef}
	tripleConstraints: ShapeExpr \fun \finset DirectedTripleConstraint
\where
	tripleConstraints(emptyshape) = \emptyset
\also
	\forall dtc: DirectedTripleConstraint; c: Cardinality @ \\
\t1		tripleConstraints(triple(dtc,c)) = \{dtc\}
\also
	\forall Exprs: \seq_1 ShapeExpr @ \\
\t1		tripleConstraints(someOf(Exprs)) = \\
\t1		tripleConstraints(oneOf(Exprs)) = \\
\t1		tripleConstraints(group(Exprs)) = \\
\t2			\bigcup \{~ Expr: \ran Exprs @ tripleConstraints(Expr) ~\}
\also
	\forall Expr: ShapeExpr; c: Cardinality @ \\
\t1		tripleConstraints(repetition(Expr,c)) = tripleConstraints(Expr)
\end{axdef}

\subsubsection{$NegatedShapeLabel$}
The semantics draft states:
\begin{quote}
for all negated shape label !T, if !T $\in$ t(n), then t1 is not a valid typing, where t1 is the typing that agrees with t everywhere, except for T $\in$ t1(n)
\end{quote}
\begin{schema}{NegatedShapeLabel}
	ValidTypingNodeLabel
\where
	negate(T) \in t(n)
\end{schema}
\begin{itemize}
\item The shape $T$ is negated at node $n$.
\end{itemize}

\subsubsection{$AssertShape$}
\begin{schema}{AssertShape}
	NegatedShapeLabel \\
	t1: Typing
\where
	t1 = t \oplus \{ n \mapsto (t(n) \setminus \{negate(T)\} \cup \{assert(T)\}) \}
\end{schema}
\begin{itemize}
\item The typing $t1$ is the same as $t$ except that at node $n$ the shape label $T$ is asserted instead of negated.
\end{itemize}

In a valid typing if any node has a negated shape, then the related typing with this shape asserted is invalid.
\begin{zed}
	\forall AssertShape @ \\
\t1		t1 \notin validTypings(G,S)
\end{zed}
Although this condition on $t(n)$ is recursive in terms of the definition of $validTypings$, 
it is well-founded since $t1(n)$ has one fewer negated shapes than $t(n)$. 
Therefore it remains to define the meaning of $validTypings$ for typings that contain no negated shapes.

\subsubsection{$assertShape$}
Given a typing $t$, node $n$, and shape label $T$ such that $negate(T) \in t(n)$, define $assertShape(t,n,T)$ to be the typing $t1$
that is the same as $t$ except that $assert(T) \in t1(n)$.
\begin{axdef}
	assertShape: Typing \cross TERM \cross ShapeLabel \pfun Typing
\where
	assertShape = \\
\t1		\{~ AssertShape @ \\
\t2			(t, n, T) \mapsto t1 ~\}
\end{axdef}

\subsubsection{$AssertedShapeLabel$}

The semantics draft defines the meaning of valid typings $t$ by imposing several conditions that must hold for all nodes $n$ and all
asserted shape labels $assert(T) \in t(n)$.
\begin{schema}{AssertedShapeLabel}
	ValidTypingNodeLabel
\where
	assert(T) \in t(n)
\end{schema}
\begin{itemize}
\item The shape label $T$ is asserted at node $n$.
\end{itemize}

The semantics draft states that the following conditions must hold for all valid typings $t$ and all nodes $n$ such that $T$ is asserted at $n$:
\begin{quote}
for all shape label $T$, if $T \in t(n)$, then there exist three mutually disjoint sets $Matching, OpenProp, Rest$ such that
\begin{enumerate}
\item $out(G, n) \cup inc(G, n) = Matching \cup OpenProp \cup Rest$, and
\item $Rest = Rest_{out} \cup Rest_{inc}$, where 
\begin{itemize}
\item $Rest_{out} = \{(out, n, p, u) \in out(G, n) | p \notin properties(expr(T, S))\}$, and 
\item $Rest_{inc} = \{(inc, u, p, n) \in inc(G, n) | p \notin invproperties(expr(T, S))\}$, and
\end{itemize}
\item $Matching$ is the union of the sets $Matching(n, t, X)$ for all triple constraint or inverse triple constraint $X$ that appears in $expr(T, S)$, and
\item if $T$ is a closed shape, then $Rest_{out} = \emptyset$ and $OpenProp = \emptyset$
\item if $T$ is an open shape, then $OpenProp \subseteq \{(out, n, p, u) \in out(G, n) | p \in incl(T, S)\}$
\item there exists a proof tree with corresponding witness mapping $wm$ for the fact that $Matching$ satisfies $expr(T, S)$, and s.t.
\begin{itemize}
\item for all outgoing triple $(out, n, p, u)$, it holds $(out, n, p, u) \in Matching(n, t, wm((out, n, p, u)))$, and moreover if $wm((out, n, p, u))$ is a shape triple constraint, then there is no value triple constraint p::C in $expr(T, S) s.t. (out, n, p, u) \in Matching(n, t, p::C)$, and
\item for all incoming triple $(inc, u, p, n) \in G$, it holds $(inc, u, p, n) \in Matching(n, t, wm((inc, u, p, n)))$, and
\item for all node $r$ that corresponds to an application of rule-one-of in the proof tree, there does not exist a valid typing $t1$ of $G$ by $S_{ri}$ s.t. $T \in t1(n)$, and
\end{itemize}
\item for all extension condition $(lang, cond)$, associated with the type $T$, $f_{lang}(G, n, cond)$ returns true or undefined.
\end{enumerate}
\end{quote}

\subsubsection{$MatchingOpenRest$}
\begin{quote}
for all shape label $T$, if $T \in t(n)$, then there exist three mutually disjoint sets $Matching, OpenProp, Rest$
\end{quote}
\begin{schema}{MatchingOpenRest}
	AssertedShapeLabel \\
	MatchingNeigh, OpenProp, Rest: \finset LabelledTriple
\where
	\disjoint \langle MatchingNeigh, OpenProp, Rest \rangle
\end{schema}
\begin{itemize}
\item There are three mutually disjoint sets of labelled triples.
\item Note that the name $MatchingNeigh$ is used to avoid conflict with the previously defined $Matching$ function.
\end{itemize}

\begin{zed}
	\forall AssertedShapeLabel @ \\
\t1		\exists MatchingNeigh, OpenProp, Rest: \finset LabelledTriple @ \\
\t2			MatchingOpenRest
\end{zed}

\subsubsection{$PartitionNeigh$}
\begin{quote}
$out(G, n) \cup inc(G, n) = Matching \cup OpenProp \cup Rest$
\end{quote}
\begin{schema}{PartitionNeigh}
	MatchingOpenRest
\where
	\langle MatchingNeigh, OpenProp, Rest \rangle \partition \\
\t1		outNeigh(G,n) \cup incNeigh(G,n)
\end{schema}

\begin{zed}
	\forall AssertedShapeLabel @ \\
\t1		\exists MatchingNeigh, OpenProp, Rest: \finset LabelledTriple @ \\
\t2			PartitionNeigh
\end{zed}

\subsubsection{$RestDef$}
\begin{quote}
$Rest = Rest_{out} \cup Rest_{inc}$, where 
\begin{itemize}
\item $Rest_{out} = \{(out, n, p, u) \in out(G, n) | p \notin properties(expr(T, S))\}$, and 
\item $Rest_{inc} = \{(inc, u, p, n) \in inc(G, n) | p \notin invproperties(expr(T, S))\}$, and
\end{itemize}
\end{quote}
\begin{schema}{RestDef}
	MatchingOpenRest \\
	Rest\_out, Rest\_inc : \finset LabelledTriple
\where
	Rest = Rest\_out \cup Rest\_inc
\also
	Rest\_out = \\
\t1		\{~ p, u: TERM | \\
\t2			out(n,p,u) \in outNeigh(G,n) \land \\
\t2			p \notin properties(expr(T,S)) @ \\
\t3				out(n,p,u) ~\}
\also
	Rest\_inc = \\
\t1		\{~ p, u: TERM | \\
\t2			inc(u,p,n) \in incNeigh(G,n) \land \\
\t2			p \notin invproperties(expr(T,S)) @\\
\t3				inc(u,p,n) ~\}
\end{schema}

\begin{zed}
	\forall MatchingOpenRest @ \\
\t1		\exists_1 Rest\_out, Rest\_inc : \finset LabelledTriple @ \\
\t2			RestDef
\end{zed}

\subsubsection{$MatchingDef$}
\begin{quote}
$Matching$ is the union of the sets $Matching(n, t, X)$ for all triple constraint or inverse triple constraint $X$ that appears in $expr(T, S)$
\end{quote}
\begin{schema}{MatchingDef}
	MatchingOpenRest 
\where
	MatchingNeigh = \\
\t1		\bigcup \{~ X:  Xs @ Matching(G, n, t, X) ~\}
\end{schema}

\begin{zed}
	\forall MatchingOpenRest @ \\
\t1		MatchingDef
\end{zed}

\subsubsection{$ClosedShapes$}
\begin{quote}
if $T$ is a closed shape, then $Rest_{out} = \emptyset$ and $OpenProp = \emptyset$
\end{quote}
\begin{schema}{ClosedShapes}
	RestDef
\where
	defT \in \ran close \implies \\
\t1		Rest\_out = \emptyset \land \\
\t1		OpenProp = \emptyset
\end{schema}

\begin{zed}
	\forall RestDef @ \\
\t1		ClosedShapes
\end{zed}

\subsubsection{$OpenShapes$}
\begin{quote}
if $T$ is an open shape, then 
\[
OpenProp \subseteq \{(out, n, p, u) \in out(G, n) | p \in incl(T, S)\}
\]
\end{quote}
\begin{schema}{OpenShapes}
	MatchingOpenRest \\
\where
	defT \in \ran open \implies \\
\t1		OpenProp \subseteq \\
\t2			\{~ p, u: TERM | \\
\t3				out(n,p,u) \in outNeigh(G,n) \land \\
\t3				p \in incl(T,S) @ \\
\t4					out(n,p,u) ~\}
\end{schema}

\begin{zed}
	\forall MatchingOpenRest @ \\
\t1		OpenShapes
\end{zed}

\subsubsection{$ProofWitness$}
\begin{quote}
there exists a proof tree with corresponding witness mapping $wm$ for the fact that $Matching$ satisfies $expr(T, S)$, and s.t.
\begin{itemize}
\item for all outgoing triple $(out, n, p, u)$, it holds $(out, n, p, u) \in Matching(n, t, wm((out, n, p, u)))$, and moreover if $wm((out, n, p, u))$ is a shape triple constraint, then there is no value triple constraint p::C in $expr(T, S) s.t. (out, n, p, u) \in Matching(n, t, p::C)$, and
\item for all incoming triple $(inc, u, p, n) \in G$, it holds $(inc, u, p, n) \in Matching(n, t, wm((inc, u, p, n)))$, and
\item for all node $r$ that corresponds to an application of rule-one-of in the proof tree, there does not exist a valid typing $t1$ of $G$ by $S_{ri}$ s.t. $T \in t1(n)$, and
\end{itemize}
\end{quote}
\begin{schema}{ProofWitness}
	MatchingDef \\
	tree: ProofTree \\
	wm: WitnessMapping
\where
	baseNeigh(tree) = MatchingNeigh \\
	baseExpr(tree) = Expr
\also
	wm = witness(tree)
\end{schema}

\begin{zed}
	\forall MatchingDef @ \\
\t1		\exists tree: ProofTree; wm: WitnessMapping @ \\
\t2			ProofWitness
\end{zed}

\subsubsection{$OutgoingTriples$}
\begin{quote}
for all outgoing triple $(out, n, p, u)$, it holds 
\[
(out, n, p, u) \in Matching(n, t, wm((out, n, p, u))), 
\]
and moreover if $wm((out, n, p, u))$ is a shape triple constraint, then there is no value triple constraint p::C in $expr(T, S)$ s.t. 
\[
(out, n, p, u) \in Matching(n, t, p::C)
\]
\end{quote}
\begin{schema}{OutgoingTriples}
	ProofWitness
\where
	\forall triple: outNeigh(G,n); p, u: TERM | \\
\t1		triple = out(n,p,u) @ \\
\t2			\LET X == wm(triple) @ \\
\t3				triple \in Matching(G, n, t, X) \land \\
\t3				(constrDTC(X) \in ShapeConstr \implies \\
\t4					\lnot (\exists C: ValueConstr | (nop(p),C) \in Xs @ \\
\t5						triple \in Matching(G, n, t, (nop(p),C))))
\end{schema}

\begin{zed}
	\forall ProofWitness @ \\
\t1		OutgoingTriples
\end{zed}

\subsubsection{$IncomingTriples$}
\begin{quote}
for all incoming triple $(inc, u, p, n) \in G$, it holds
\[
(inc, u, p, n) \in Matching(n, t, wm((inc, u, p, n)))
\]
\end{quote}
\begin{schema}{IncomingTriples}
	ProofWitness
\where
	\forall triple: incNeigh(G,n) @ \\
\t1		\LET X == wm(triple) @ \\
\t2			triple \in Matching(G, n, t, X)
\end{schema}

\begin{zed}
	\forall ProofWitness @ \\
\t1		IncomingTriples
\end{zed}

\subsubsection{$OneOfNodes$}
\begin{quote}
for all node $r$ that corresponds to an application of rule-one-of in the proof tree, 
there does not exist a valid typing $t1$ of $G$ by $S_{ri}$ s.t. $T \in t1(n)$
\end{quote}

Let $OneOfNodes$ describe the situation where we are
given a graph $G$, a schema $S$, a typing $t$ of $G$ under $S$, a node $n$ in $G$, a shape label $T$
in $t(n)$, a proof tree $tree$ for the triples $MatchNeigh$ and the expression $Expr= expr(T,S)$ and an application
of rule-one-of $r$ in the proof tree.

\begin{schema}{OneOfNodes}
	ProofWitness \\
	RuleOneOfApplication \\
	Expr\_ri : ShapeExpr \\
	S\_ri : SchemaWD
\where
	Expr\_ri = reduceExpr(tree, path)
\also
	S\_ri = replaceShape(S, T, Expr\_ri)
\end{schema}

Whenever rule-one-of is applied in the proof tree, there must not be any valid typings $t1$
for the reduced schema $S\_ri$ in which the selected component of the one-of shape expression is eliminated.

\begin{zed}
	\forall OneOfNodes @ \\
\t1		\lnot (\exists t1: validTypings(G,S\_ri) @ \\
\t2			assert(T) \in t1(n))
\end{zed}

\subsubsection{$ExtensionConditions$}
\begin{quote}
for all extension condition $(lang, cond)$, associated with the type $T$, $f_{lang}(G, n, cond)$ returns true or undefined
\end{quote}

The semantics of an extension condition is given by a language oracle function
that evaluates the extension condition $cond$ on a pointed graph $(G,n)$
and returns a code indicating whether the pointed graph
satisfies the extension condition, or if an error condition holds, or if the extension condition is undefined.
\begin{axdef}
	f : ExtLangName \cross Graph \cross TERM \cross ExtDefinition \fun ReturnCode
\where
	\forall G: Graph; n: TERM| (G, n) \in PointedGraph @ \\
\t1		\forall lang: ExtLangName; cond: ExtDefinition @ \\
\t2			\LET returnCode == f(lang, G, n, cond) @ \\
\t3				returnCode = trueRC \implies (G,n) \notin violatedBy(lang, cond) \land \\
\t3				returnCode = falseRC \implies (G,n) \in violatedBy(lang, cond)
\end{axdef}
\begin{itemize}
\item If the oracle returns true then the pointed graph satisfies the extension condition.
\item If the oracle returns false then the pointed graph violates the extension condition.
\end{itemize}

Let the return codes for the language oracles be $ReturnCode$.
\begin{zed}
	ReturnCode ::= trueRC | falseRC | errorRC | undefinedRC
\end{zed}
\begin{itemize}
\item true means the extension condition is satisfied.
\item false means the extension condition is violated.
\item error means an error occurred.
\item undefined means the extension condition is undefined.
\end{itemize}

\begin{schema}{ExtensionConditions}
	MatchingOpenRest \\
	lang: ExtLangName \\
	cond: ExtDefinition
\where
	\LET ecs == extConds(ruleT) @ \\
\t1		(lang,cond) \in \ran ecs
\end{schema}
\begin{itemize}
\item $(lang, cond)$ is an extension condition for $T$.
\end{itemize}

\begin{zed}
	\forall ExtensionConditions @ \\
\t1		f(lang, G, n, cond) \in \{trueRC, undefinedRC \}
\end{zed}

\section{Issues}
\label{sec-issues}

Some areas of the semantics draft have multiple interpretations or appear to be wrong and therefore require further clarification.
These areas are discussed below.

\subsection{{\tt dep-subgraph(T,S)}}
In the definition of {\tt dep-subgraph(T,S)}, is the shape {\tt T} considered to be reachable from itself?

\subsection{{\tt negshapes(S)}}
In the definition of {\tt negshapes(S)}, the third bullet states:

\begin{verbatim}
there is a shape label T1 and a shape triple constraint p::C, 
or an inverse shape triple constraints ^p::C in expr(T1, S), 
and T appears in C.
\end{verbatim}

This statement looks wrong because it omits mention of negation. 
If there is no negation involved, why would {\tt T} be in {\tt negshapes(S)}?

Does this definition only select directed triple constraints that have cardinality {\tt [1,1]}
because that is the default?
If not then {\tt negshapes(S)} is the set of all labels that are referenced in any shape definition ($refs(S)$), which seems wrong.

\subsection{Triple matches constraint}
The definition of matching {\tt p:C} and {\tt \verb+^+p:C} omits consideration of {\tt C}.
The explanation is as follows.

\begin{verbatim}
The following definition introduces the notion of satisfiability 
of a shape constraint by a set of triples. Such satisfiability 
is going to be used for checking that the neighborhood of 
a node satisfies locally the constraints defined by a shape 
expression, without taking into account whether the shapes 
required by the triple constraints and inverse triple constraints 
are satisfied.
\end{verbatim}

This statement implies that only shape constraints should be ignored here.
However, the definition ignores the value set constraints too.
This looks wrong.

\subsection{{\tt rule-triple-constraint}}
Add the condition that all the outgoing triples must be distinct.

\subsection{{\tt rule-inverse-triple-constraint}}
Add the condition that all the incoming triples must be distinct.

\subsection{{\tt rule-group}}
Add the condition that {\tt i} and {\tt j} must be different.

\subsection{{\tt rule-repeat}}
Add the condition that {\tt i} and {\tt j} must be different.

\subsection{Reduced Schema for {\tt rule-one-of}}
This is an edge case.
It only makes sense to reduce the schema if there are more than one components.
Applying {\tt rule-one-of} to a sequence of one shape is equivalent to requiring that shape.
Add this condition to the definition.

\subsection{Reduced Schema for {\tt rule-one-of} under a repetition expression}
Something looks wrong here because if a repetition expression has a one-of expression as a child then there is no way
to associate the reduced one-of expression with just the path taken in the proof tree since all the children of a repetition expression
share the same shape expression. However, a rule-repeat node in the proof tree has many children and there is no requirement
that all children would use the same branch of the one-of expression.

\section{Conclusion}
\label{sec-conclusion}

The exercise of formalizing the semantics draft has resulted in a considerable expansion in the size of the document.
The result has been the identification of a number of quality issues.
This exercise has also established that the recursive definitions in the semantics draft are well-founded.
However, it is not clear that these definitions produce results that agree with our intuition, or that they can be computed efficiently.

One possible way to further validate the semantics draft is to translate it into an executable formal specification system
such as Coq \cite{bertot:coq} and test it on a set of examples, including both typical documents and corner cases.

\bibliography{z-core-shacl-semantics}
\end{document}